\documentclass[12pt, draftclsnofoot, onecolumn]{IEEEtran}

\usepackage{graphicx}
\usepackage{enumerate}
\usepackage{cite}
\usepackage{bm}
\usepackage[dvipsnames]{xcolor}
\usepackage{amsmath,amssymb,amsthm}
\usepackage{dutchcal}
\usepackage{stfloats}
\usepackage{suffix}
\usepackage{mathtools}

\DeclareMathAlphabet{\mathpzc}{OT1}{pzc}{m}{it}

\renewcommand*{\arraystretch}{1.5}
\newcommand{\kron}{ \hspace{-.8mm} \otimes \hspace{-.8mm}}
\newcommand{\kr}{ \hspace{-.8mm} \odot \hspace{-.8mm}}

\usepackage[font=footnotesize,skip=2pt]{caption}
\DeclarePairedDelimiterX\MeijerM[3]{\lparen}{\rparen}%
{\begin{smallmatrix}#1 \\ #2\end{smallmatrix}\delimsize\vert\,#3}

\newcommand\MeijerG[8][]{%
  G^{\,#2,#3}_{#4,#5}\MeijerM[#1]{#6}{#7}{#8}}

\WithSuffix\newcommand\MeijerG*[7]{%
  G^{\,#1,#2}_{#3,#4}\MeijerM*{#5}{#6}{#7}}

\begin{document}

\title{IRS-Assisted Massive MIMO-NOMA Networks: Exploiting Wave Polarization \vspace{-3mm}}

\author{Arthur S. de Sena, \textit{Student Member}, \textit{IEEE}, Pedro H. J. Nardelli,  \textit{Senior Member}, \textit{IEEE}, Daniel B. da Costa, \textit{Senior Member}, \textit{IEEE}, F. Rafael M. Lima, \textit{Senior Member}, \textit{IEEE}, Liang Yang, \textit{Member}, \textit{IEEE}, Petar Popovski, \textit{Fellow}, \textit{IEEE},
Zhiguo Ding, \textit{Fellow}, \textit{IEEE}, and Constantinos B. Papadias, \textit{Fellow}, \textit{IEEE}

\thanks{A. S. de Sena and Pedro H. J. Nardelli are with Lappeenranta-Lahti University of Technology, Finland, (email: arthurssena@ieee.org, pedro.nardelli@lut.fi).}
\thanks{D. B. da Costa and F. Rafael M. Lima are with the Federal University of Cear\'{a}, Brazil (email: danielbcosta@ieee.org, rafaelm@gtel.ufc.br).}
\thanks{L. Yang is with Hunan University, China (email: liangy@hnu.edu.cn).}
\thanks{P. Popovski is with Aalborg University, Denmark (email: petarp@es.aau.dk).}
\thanks{Z. Ding is with the University of Manchester, UK (email: zhiguo.ding@manchester.ac.uk).}
\thanks{C. B. Papadias is with the American College of Greece, Greece (email: cpapadias@acg.edu).}
% <-this % stops a space
}

\maketitle

\vspace{-20mm}
\begin{abstract}
A dual-polarized intelligent reflecting surface (IRS) can contribute to a better multiplexing of interfering wireless users. In this paper, we use this feature to improve the performance of dual-polarized massive multiple-input multiple-output (MIMO) with non-orthogonal multiple access (NOMA) under imperfect successive interference cancellation (SIC). By considering the downlink of a multi-cluster scenario, the IRSs assist the base station (BS) to multiplex subsets of users in the polarization domain. Our novel strategy alleviates the impact of imperfect SIC and enables users to exploit polarization diversity with near-zero inter-subset interference. To this end, the IRSs are optimized to mitigate transmissions originated at the BS from the interfering polarization. The formulated optimization is transformed into quadratic constrained quadratic sub-problems, which makes it possible to obtain the optimal solution via interior-points methods. We also derive analytically a closed-form expression for the users' ergodic rates by considering large numbers of reflecting elements. This is followed by representative simulation examples and comprehensive discussions. The results show that when the IRSs are large enough, the proposed scheme always outperforms conventional massive MIMO-NOMA and MIMO-OMA systems even if SIC error propagation is present. It is also confirmed that dual-polarized IRSs can make cross-polar transmissions beneficial to the users, allowing them to improve their performance through diversity.

\end{abstract}\vspace{-4mm}

\begin{IEEEkeywords}\vspace{-2mm}
	Multi-polarization, intelligent reflecting surfaces, Massive MIMO, NOMA
\end{IEEEkeywords}

\IEEEpeerreviewmaketitle

\section{Introduction}

The fifth-generation (5G) wireless systems are already being deployed worldwide. The novel technologies and infrastructures of 5G provide support to unprecedented applications with diverse requirements, such as high data rates, high reliability, and low latency. One key technology is massive multiple-input multiple-output (MIMO), where a large number of antennas at the base station (BS) is used to transmit parallel data streams to multiple users through spatially separated beams. Conventionally, orthogonal multiple access (OMA) techniques are combined with massive MIMO to guarantee zero inter-beam interference in scenarios where it is difficult to multiplex users solely in the space domain. Even though such schemes can effectively cope with the interference issue, they may perform poorly in terms of spectral efficiency and latency as the number of users increases. Therefore, MIMO-OMA systems are not ideal for ultra-dense deployments, and this motivates the use of non-orthogonal multiple access (NOMA), such that MIMO-NOMA can serve simultaneously several users with non-separable beams.

The performance of a massive MIMO-NOMA network scales up with the increase of transmit and receive antennas. However, due to physical space constraints, the number of antennas installed in practical systems is limited at both the BS and user's devices. One efficient strategy to alleviate such a limitation can be achieved by arranging the antenna elements into co-located pairs with orthogonal polarizations, forming a dual-polarized antenna array. With such an approach, it becomes possible to install twice the number of antennas of a single-polarized array utilizing the same physical space. In addition, since antennas with orthogonal polarizations exhibit a low correlation, dual-polarization enables massive MIMO-NOMA systems to exploit diversity in the polarization domain, which can significantly outperform conventional single-polarized schemes \cite{ni3}. Due to these attractive features, dual-polarized antenna arrays have been adopted as standard in the 3rd generation partnership project (3GPP) long-term evolution advanced (LTE-A) and 5G New Radio (NR) specifications \cite{Asplund2020}.

Despite the mentioned advantages, a dual-polarized massive MIMO-NOMA system still has numerous limitations. For instance, the mutual coupling between antennas and the stochastic nature of the scatterer environment can depolarize the transmitted signals and generate cross-polar interference at the receivers. As demonstrated in \cite{ni3}, these depolarization phenomena can deteriorate the system performance. Furthermore, in power-domain NOMA, the users need to employ successive interference cancellation (SIC) to decode their received data symbols, which also has some drawbacks. An increase in the number of users leads to higher interference and a more complex SIC decoding process, potentially resulting in excessive decoding errors, lowered system throughput, and increased usage of the device battery. It was shown in~\cite{SenaISIC2020} that SIC errors severely impact the performance of massive MIMO-NOMA systems, making them less spectrally efficient than massive MIMO-OMA schemes. This harmful characteristic limits the maximum number of users served with NOMA in practical systems.

This implies that the benefits of dual-polarized MIMO-NOMA systems can be harvested if there is an increased control of the (de)polarization properties of the propagation environment. In this sense, the recent concept of an intelligent reflecting surface (IRS)~\cite{Renzo2020} holds a great potential. An IRS is an engineered device that comprises multiple sub-wavelength reflecting elements with reconfigurable electromagnetic properties. The phases and amplitudes of reflections induced by the IRS elements are controlled independently via software, which enables them to, collectively, forward the impinging waves with an optimized radiation pattern and reach diverse objectives like beam steering, collimation, absorption, and control of polarization \cite{aswc2020}. Such appealing features unlock countless new possibilities for manipulating the random phenomena of electromagnetic propagation, a critical issue in any wireless communication system. This is discussed in several recent works, some of them dealing specifically with MIMO-OMA and MIMO-NOMA. \vspace{-2mm}

\subsection{Related Works}
The majority of recent IRS-MIMO related works are concentrated on the study of point-to-point or OMA-based schemes. For example, the authors of \cite{pwr01} investigated the performance of IRS-assisted point-to-point narrow-band and orthogonal frequency division multiplexing (OFDM) MIMO systems. Specifically, transmit beamforming and IRS reflecting elements were optimized to maximize the ergodic rates of the considered systems. In the simulation examples, the proposed optimization algorithms outperformed conventional MIMO schemes with and without IRSs.
The minimization of the symbol error rate (SER) of an IRS-assisted point-to-point MIMO system was addressed in \cite{pwr02}. The IRS reflecting elements and beamforming matrix were optimized alternatively, in which four different methods were investigated. All methods achieved superior performance than conventional systems without IRS in terms of SER. %The proposed scheme also outperformed MIMO systems assisted by full-duplex relays.
A single-cell multi-user OMA-based network was considered in \cite{pwr03}. The authors of this work minimized the total transmit power of an IRS-MIMO system under users’ individual SINR constraints. An asymptotic analysis with a large number of reflecting elements was also performed.
%It was found through simulations that, with fewer transmit antennas, the proposed IRS-MIMO system can reach the same rate performance of a massive MIMO system without IRS.
%
The multi-cell IRS-MIMO case was addressed in \cite{pwr04}. In this work, an IRS was exploited to improve the performance of cell-edge users, in which two algorithms based on majorization-minimization and the complex circle manifold methods were proposed to optimize the IRS reflecting elements. 
The authors of \cite{pwr05} employed an IRS to assist multi-user MIMO cognitive radio systems, where a block coordinate descent algorithm was proposed to maximize the achievable weighted sum rate.
The employment of IRSs for improving the performance of simultaneous wireless information and power transfer (SWIPT) in MIMO systems was investigated in \cite{pwr06},
and for addressing security issues in \cite{pwr08}.

A few contributions have investigated IRSs in MIMO-NOMA schemes. For instance, the work in \cite{rw01} addressed a simple IRS-assisted MIMO-NOMA network, in which near and far users were paired to be served with NOMA with the aid of IRSs. %The authors carried out an analytical analysis and derived a closed-form expression for the outage probability. The performance of the proposed scheme was investigated through simulation examples, where hardware impairments were also taken into account.
The energy efficiency of a two-user IRS-MIMO-NOMA network was investigated in \cite{rw02}. In this work, the IRS reflecting elements and the beamforming vectors at the BS were jointly optimized to minimize the total power consumption of the system. %In the simulation results, the proposed system outperformed conventional MIMO-NOMA and MIMO-OMA schemes in terms of energy efficiency.
In \cite{rw03}, by considering both continuous and discrete phase shifters, the authors maximized the sum-rate of a IRS-MIMO-NOMA system in a scenario with multiple users. The proposed scheme remarkably outperformed conventional NOMA and OMA-based systems in the presented simulation examples. %Moreover, 3-bit phase shifters were enough to reach almost the same performance as the ideal case with infinity resolution.
A multi-cluster IRS-assisted MIMO-NOMA network was considered in \cite{rw05}. By relaxing the need for active beamforming at the BS, the authors focused on the design of an IRS for canceling inter-cluster interference.% Closed-form expressions for the outage probability and ergodic rates were derived. The proposed system achieved better performance than zero-forcing and signal-alignment-based schemes in all presented simulation examples.
The application of IRSs to millimeter-wave NOMA systems was studied in \cite{rw04}. With the objective of maximizing the system sum-rate, this work developed an algorithm for optimizing power allocation, reflecting elements, and active beamforming.% The obtained solution rendered significant performance gains to the proposed system, which outperformed the millimeter-wave OMA counterpart.
The scenario with IRSs mounted on unmanned aerial vehicles (UAV) to assist a MIMO-NOMA network was investigated in \cite{rw06}. In this work, by optimizing the position of the UAV, the transmit beamforming, and the IRS reflecting elements, the rate of the strong user was maximized while guaranteeing the target rate of the weak user. %Simulation results demonstrated impressive rate gains of the proposed scheme over UAV-assisted orthogonal frequency-division multiple access (OFDMA) systems.

\vspace{-2mm}
\subsection{Motivation and Contributions}

To the best of our knowledge, all related works are limited to only single-polarized systems, and there are no works that exploit the capabilities of IRSs for manipulating wave polarization in dual-polarized MIMO-NOMA networks. Motivated by this, and given the great potential of IRSs for improving the performance of communication systems, in this paper, we harness the attractive features of dual-polarized IRSs for enabling users to exploit polarization diversity and for reducing the impact of imperfect SIC in a multi-cluster dual-polarized MIMO-NOMA network. Further details and the main contributions of this work are summarized as follows:
\begin{itemize}
    \item By considering a scenario where the BS and the users employ multiple dual-polarized antennas and assuming imperfect SIC, we propose a novel strategy that exploits the functionalities of dual-polarized IRSs to assist the BS to subdivide each group of users into two polarization subsets. For users in the first subset, the BS transmits the data symbols using vertically polarized antennas and, for users in the second one, the BS transmits using the horizontally polarized antennas. With this strategy, SIC can be executed by users from each subset separately. As a result, each user will experience less SIC interference when decoding its message. Moreover, the IRSs transform depolarization phenomena into an advantage and enable the users to exploit polarization diversity with near-zero inter-subset interference.
    
    \item By assuming that the users and the IRSs are distributed among different spatial clusters and aiming to focus the transmissions to the users and IRSs of interest and null out anywhere else, we first exploit the second-order statistics of the channels, i.e., the channel covariance matrices, to construct the active beamforming matrices at the BS. We then concatenate the beamforming matrix for spatial interference cancellation with a low-complexity precoding vector that is designed to multiplex the users and form the polarization subsets. 
    
    \item The dual-polarized reflecting elements of each IRS are optimized to mitigate the transmissions originated at the BS from the interfering polarization. The formulated optimization problem is challenging to solve. To overcome the complex formulation, we transform the original problem into quadratic constrained quadratic sub-problems, and we show that their optimal solutions can be obtained via interior-points methods in polynomial time.
    
    \item An in-depth performance analysis is carried out, where, by modeling polarization interference and errors from imperfect SIC, we derive the signal-to-interference-plus-noise ratio (SINR) experienced by the users and investigate the statistical distributions of the effective channel gains. Because the reflecting elements of the IRSs change rapidly with the fast fading channels, identifying the exact distributions for arbitrary numbers of reflecting elements becomes difficult. As an alternative, we characterize the approximate distributions for the asymptotic case with a large number of reflecting elements. Based on this asymptotic statistical analysis, we derive a closed-form expression for the ergodic rates observed by each user, which provides a practical tool for verifying the fundamental limits of the proposed system when large IRSs are employed.
    
    \item Last, by presenting representative numerical simulation results, we validate the analysis and supplement it with discussions. We show that when the IRSs are large enough, the proposed scheme always outperforms conventional massive MIMO-NOMA and MIMO-OMA systems even if SIC error propagation is present. We also confirm that the dual-polarized IRSs can make cross-polar transmissions beneficial to the users, allowing them to improve their performance through diversity.

\end{itemize}

\noindent  {\bf Notation and Special Functions:} Bold-faced lower-case letters denote vectors and upper-case represent matrices. The $i$th element of a vector $\mathbf{a}$ is denoted by $[\mathbf{a}]_i$, the $(ij)$ entry of a matrix $\mathbf{A}$ by $[\mathbf{A}]_{ij}$, and the transpose and the Hermitian transpose of $\mathbf{A}$ are represented by $\mathbf{A}^T$ and $\mathbf{A}^H$, respectively. The symbol $\otimes$ represents the Kronecker product, $\odot$ is the Khatri-Rao product \cite{Brewer78}, $\mathbf{I}_M$ represents the identity matrix of dimension $M\times M$, and $\mathbf{0}_{M, N}$ denotes the $M\times N$ matrix with all zero entries. The operator $\textit{vec}\{\cdot \}$ transforms a matrix of dimension $M\times N$ into a column vector of length $MN$, the operator $\textit{vecd}\{\cdot \}$ converts the diagonal elements of an $M\times M$ square matrix into a column vector of length $M$, and $\textit{diag}\{\cdot \}$ transforms a vector of length $M$ into an $M\times M$ diagonal matrix. In addition, $\Re\{\cdot \}$ returns the real part of a complex number, $(\cdot)^{*}$ is the complex conjugate, $\mathbb{E}[\cdot]$ denotes expectation, $\Gamma(\cdot)$ is the Gamma function \cite[eq. (8.310.1)]{ref8}, $\gamma(\cdot,\cdot)$ is the lower incomplete Gamma function \cite[eq. (8.350.1)]{ref8}, and $\MeijerG*{m}{n}{p}{q}{\mathbf{a}}{ \mathbf{b}}{x}$ corresponds to the Meijer's G-function \cite[eq. (9.301)]{ref8}. \vspace{-2mm}

\section{Fundamentals of a dual-polarized IRS}\label{secfund}
\begin{figure}
	\centering
	\includegraphics[width=.8\linewidth]{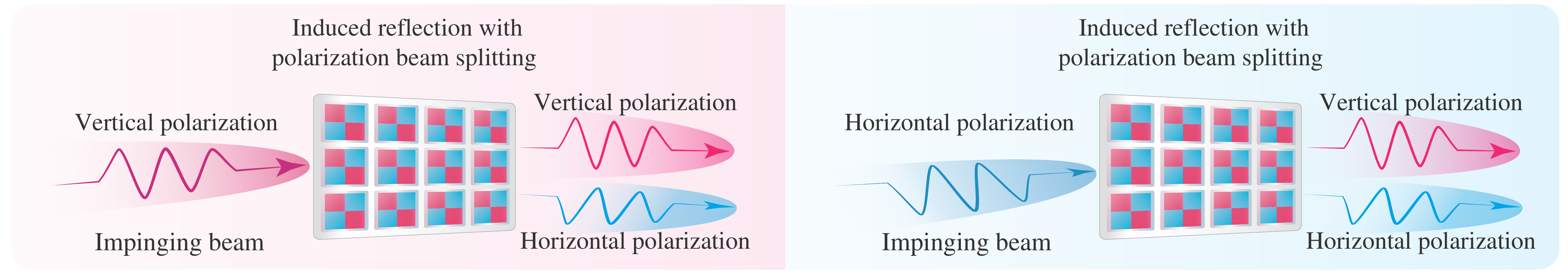}
	\caption{Simplified capabilities of a dual-polarized IRS. Linearly polarized impinging signals are split into two beams with orthogonal polarizations.}\label{fig:irs_c}
\end{figure}

The design of dual-polarized IRSs and their potential capabilities have been well studied in the field of antennas and electromagnetic theory \cite{Sun18,JWang2020, Ma2015}. In addition to phase/amplitude control, also possible with a single-polarized IRS, a dual-polarized IRS can perform polarization beam splitting, independent control of impinging polarizations, and polarization conversion \cite{Ma2015}. For instance, by properly tuning the IRS reflecting elements, it is possible to convert a vertically polarized wave into a horizontally polarized one, and vice-versa, or reflect it with its original polarization \cite{Sun18,JWang2020}. These features can find useful applications in dual-polarized communication systems,  such as interference mitigation or polarization diversity. Specifically, by considering linear vertical-horizontal polarization, the transformations induced by each reflecting element of a dual-polarized IRS can be modeled by a reflection matrix:
\begin{align}\label{refmatsing}
    \mathbf{\Psi} = \renewcommand*{\arraystretch}{1}\begin{bmatrix}
            \omega^{vv}e^{-j\phi^{vv}} & \omega^{hv}e^{-j\phi^{hv}} \\
            \omega^{vh}e^{-j\phi^{vh}} & \omega^{hh}e^{-j\phi^{hh}}
    \end{bmatrix},
\end{align}
where $\phi^{pq}_{l} \in [0,2\pi]$ and $\omega^{pq}_{l} \in [0,1]$ represent, respectively, the phase and amplitude of reflection induced by the IRS element from polarization $p$ to polarization $q$, with $p,q \in \{v,h\}$, in which $v$ stands for vertical and $h$ for horizontal. A simplified illustration of the capabilities of the dual-polarized IRS considered in this work is shown in Fig. \ref{fig:irs_c}.

We illustrate these concepts through a simple example. Suppose that a transmitter that is equipped with a single vertically polarized antenna sends information to a receiver that employs a pair of dual-polarized antennas, with a vertically and a horizontally polarized antenna element, respectively. In an ideal scenario without any depolarization, the transmitted information would only be received in the matching vertically polarized receive antenna, becoming impossible to explore polarization diversity at the receiver. By deploying a dual-polarized IRS, the transmitted vertically polarized wave can be split into two independent beams, one with vertical polarization and another with horizontal polarization, as shown in Fig. \ref{fig:irs_c}. This would enable the receiver to exploit polarization diversity and improve its performance. 
Specifically, assume that the IRS has only a single dual-polarized reflecting element. Then, by recalling the dyadic backscatter channel model \cite{Liang2019,Renzo2020}, and using the reflection matrix in \eqref{refmatsing}, the noiseless signal propagated through the reflected IRS link, observed at the vertically and horizontally polarized receive antennas is:
\begin{align}\label{expi}\setlength{\arraycolsep}{5pt} \renewcommand*{\arraystretch}{1}
    \begin{bmatrix}
        y^v \\
        y^h
    \end{bmatrix} \hspace{-1.5mm}  = \frac{1}{\sqrt{2}}
    \hspace{-1.5mm} 
    \begin{bmatrix}
            (s^{vv})^{*} & 0 \\
            0 & (s^{hh})^{*}
    \end{bmatrix} 
    \begin{bmatrix}
            \omega^{vv}e^{-j\phi^{vv}} & \omega^{hv}e^{-j\phi^{hv}} \\
            \omega^{vh}e^{-j\phi^{vh}} & \omega^{hh}e^{-j\phi^{hh}}
    \end{bmatrix}
    \hspace{-2mm}
    \begin{bmatrix}
            g^{vv} \\
            0
    \end{bmatrix} \hspace{-1.5mm} x = \hspace{-1.5mm}\begin{bmatrix}
            \frac{1}{\sqrt{2}}(s^{vv})^{*}\omega^{vv}e^{-j\phi^{vv}} g^{vv} x \\ 
            \frac{1}{\sqrt{2}}(s^{hh})^{*}\omega^{vh}e^{-j\phi^{vh}} g^{vv} x
    \end{bmatrix}\hspace{-1.5mm},
\end{align}
where $x$ is the transmitted data symbol, $g^{vv}$ is the channel coefficient between the transmitter and the IRS, and $s^{pq}$ is the channel coefficient between the IRS and the receiver corresponding to the signal that was reflected with polarization $p$ and arrived with polarization $q$, in which $p, q \in \{v,h\}$. Since an IRS is a passive device, we introduce a normalization factor of $\frac{1}{\sqrt{2}}$. As one can observe in \eqref{expi}, by performing polarization beam splitting, the IRS was capable of delivering two replicas with independent phases and amplitudes of the transmitted data symbol. Hence, a range of new possibilities can be enabled by properly optimizing the IRS reflecting elements. For instance, if the transmitter in this example is instead sending interference, one could easily switch the IRS to an absorption mode, i.e., set the coefficients $\omega^{vv}$ and $\omega^{vh}$ to zero, so that no interfering transmissions would arrive at the receiver.

In this work, we consider a generalization of the signal model in \eqref{expi} in a more complex setup containing several IRSs with a large number of dual-polarized reflecting elements. Despite the more complex system model and the greater number of reflecting elements, the capabilities of the larger IRSs considered in the proposed scheme are the same as the presented in this section, i.e., capabilities of manipulating wave polarization, clearly illustrated in Fig. \ref{fig:irs_c}. Since the basic background for understanding this work's proposal has been provided, we can now dive into the detailed system model. \vspace{-2mm}

\section{System Model}

\begin{figure}
	\centering
	\includegraphics[width=.5\linewidth]{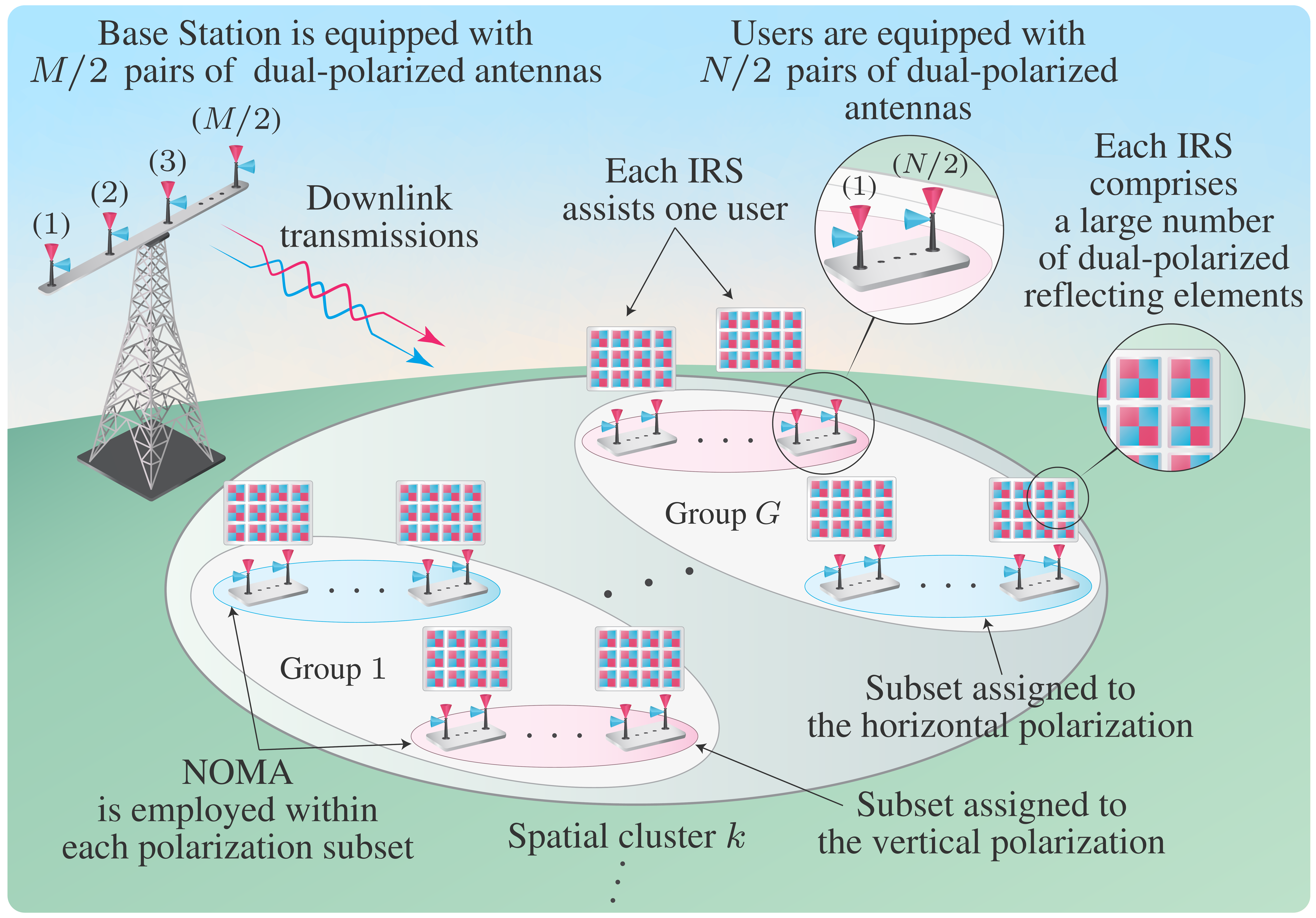}
	\caption{System model. Dual-polarized IRSs enable users to exploit polarization diversity by mitigating polarization interference.}\label{fig:sysmodel}
\end{figure}

\begin{figure}
	\centering
	\includegraphics[width=.8\linewidth]{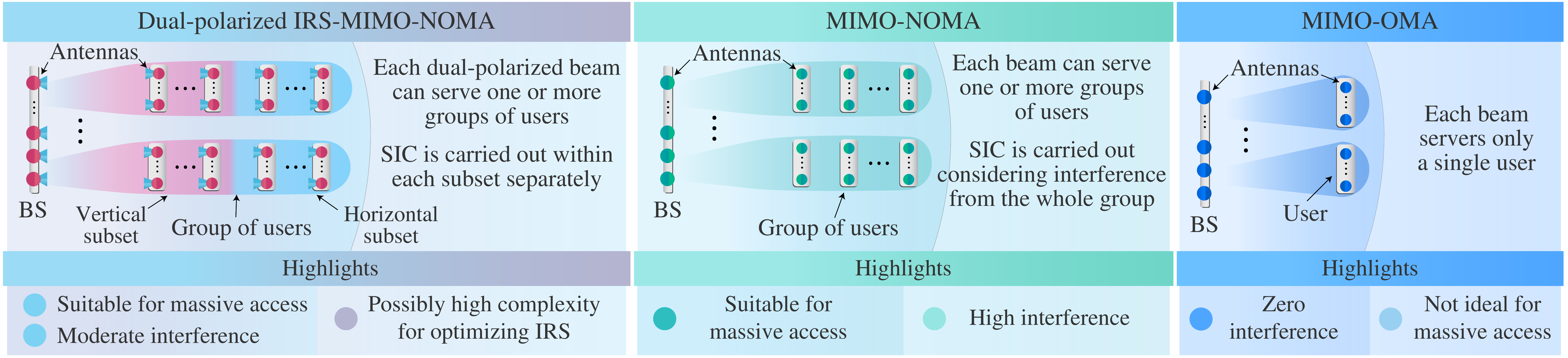}
	\caption{Main differences between the proposed dual-polarized IRS-MIMO-NOMA and other conventional schemes.}\label{fig:comp}
\end{figure}

Consider a single cell MIMO-NOMA network where a single BS is communicating in downlink mode with multiple users. Both users and the BS comprise dual-polarized antenna elements that are arranged into multiple co-located pairs, each one containing one vertically and one horizontally polarized antenna element. More specifically, users are equipped with $N/2$ pairs of dual-polarized receive antennas, and the BS with $M/2$ pairs of dual-polarized transmit antennas that are organized in a uniform linear array. It is considered that $M$ and $N$ are even, and that $M\gg N$. Moreover, within the cell, users are assumed to be distributed among different geographical areas, forming $K$ spatial clusters with $Q$ users each. Users within each cluster are organized into $G$ groups, each one containing $U$ users, i.e, $Q = GU$. In conventional MIMO-NOMA systems, the $u$th user from a given group performs SIC by considering interference from all the other $U-1$ users within the same group. However, since SIC is an interference-limited technique, such an approach can lead to performance degradation, which here is tackled by a novel strategy that exploits the polarization domain. Specifically, we program the BS to further subdivide each of the $G$ groups into two polarization subsets, namely vertical subset and horizontal subset, each one containing $U^{p}$ users, $p\in \{v,h\}$, i.e., $U^{v}$ users are served with vertically polarized transmit antennas, and $U^{h}$ users are served with horizontally polarized antennas, such that $U^{v} + U^{h} = U$.
To enable this scheme, we exploit the capabilities of dual-polarized IRSs to ensure that signals transmitted from one polarization impinge only at users assigned to that specific polarization. For instance, if a user is assigned to the vertical polarization, its serving IRS should cancel out all signals coming from horizontally polarized BS antennas. For this, we assume that there are $U$ IRSs with $L$ dual-polarized reflecting elements installed within each group and that each IRS assists exactly one user\footnote{In practice, more than one user can be connected simultaneously to an IRS. However, as stated in \cite{aswc2020}, as the number of connected users increases, the complexity for optimizing the IRS reflecting elements also increases. Because of this, the number of users is usually maintained small. Despite that, investigating the performance of the proposed system with multiple users connected to each IRS is also interesting, but this possibility is left for future works.}, as illustrated in Fig. \ref{fig:sysmodel}. A comparison of the main characteristics and highlights between our proposal and those from conventional schemes is provided in Fig. \ref{fig:comp}. As more details will be provided later, in addition to reducing interference and SIC decoding errors, our IRS-MIMO-NOMA scheme naturally enables users to exploit polarization diversity with only a low computational complexity.

Following the proposed strategy, after the users have been properly grouped, the BS applies superposition coding to each polarization subset and transmit the superimposed messages through the assigned polarization. More specifically, the BS sends the following signal
\begin{align}\label{eq03}
\mathbf{x} =  \sum_{k=1}^{K} \mathbf{P}_k\renewcommand*{\arraystretch}{.8} \begin{bmatrix}
\mathbf{x}^v \\ \mathbf{x}^h
\end{bmatrix} = \sum_{k=1}^{K} \mathbf{P}_k \sum_{g=1}^{G} \sum_{u=1}^{U} \mathbf{v}_{kgu} 
\alpha_{kgu}{x}_{kgu} \in \mathbb{C}^{M \times 1},
\end{align}
where $\mathbf{x}^p$ is the data vector transmitted in the polarization $p \in \{v,h\}$. ${x}_{kgu} $ and  $\alpha_{kgu}$  are, respectively, the symbol and the power coefficient for the $u$th user in the $g$th group within the $k$th cluster. $\mathbf{P}_k \in \mathbb{C}^{M\times \bar{M}}$ is a precoding matrix intended to eliminate inter-cluster interference, in which $\bar{M}$ is a parameter that controls the number of effective data streams transmitted from the BS, and $\mathbf{v}_{kgu} = \left[ (\mathbf{v}^v_{kgu})^T, (\mathbf{v}^h_{kgu})^T  \right]^T \in \mathbb{C}^{\bar{M} \times 1}$ is an inner precoding vector responsible for multiplexing the users in the polarization domain, satisfying $\|\mathbf{v}_{kgu}\|^2 = 1$.

As the phases and amplitudes of reflections induced by a single dual-polarized reflecting element can be modeled by the $2\times 2$ matrix in \eqref{refmatsing}, the reflection matrix for an IRS with $L$ reflecting elements can be generalized to a $2L\times 2L$ matrix. This matrix is partitioned into four $L\times L$ diagonal sub-matrices. Thus, the reflection matrix for the dual-polarized IRS that assists the $u$th user in the $g$th group of the $k$th spatial cluster is:
\begin{align}\label{coeffmtx}
\bm{\Theta}_{kgu} = \renewcommand*{\arraystretch}{1}
    \begin{bmatrix}
        \bm{\Phi}_{kgu}^{vv} & \bm{\Phi}_{kgu}^{hv} \\
        \bm{\Phi}_{kgu}^{vh} & \bm{\Phi}_{kgu}^{hh}
    \end{bmatrix} \in \mathbb{C}^{2L\times 2L},
\end{align}
where $\bm{\Phi}_{kgu}^{pq} = \textit{diag}\{[\omega^{pq}_{kgu,1} e^{-j\phi^{pq}_{kgu,1}},\omega^{pq}_{kgu,2} e^{-j\phi^{pq}_{kgu,2}}, \cdots,$ $\omega^{pq}_{kgu,L} e^{-j\phi^{pq}_{kgu,L}}]\} \in \mathbb{C}^{L\times L}$, with $\phi^{pq}_{kgu,l}$ and $\omega^{pq}_{kgu,l}$ representing, respectively, the phase and amplitude of reflection induced by the $l$th IRS element from polarization $p$ to polarization $q$, with $p,q \in \{v,h\}$, in which we must have $|\omega^{pq}_{kgu,l}|^2 \leq 1$ for passive reflection. By using the multi-polarized and the dyadic backscatter channel models \cite{Liang2019,ni3,Renzo2020}, the composite full dual-polarized channel matrix for the $u$th user in the $g$th group of the $k$th cluster can represented by \begin{align}\label{eq:ch0} 
    \mathbf{H}^H_{kgu} &= 
    \sqrt{\zeta_{kgu}^\text{\tiny BS-IRS} \zeta_{kgu}^\text{\tiny IRS-U}}
    \frac{1}{\sqrt{2}}\renewcommand*{\arraystretch}{1}
    \begin{bmatrix}
        \mathbf{\bar{S}}_{kgu}^{ vv} &
        \mathbf{0}_{L,\frac{N}{2}} \\
        \mathbf{0}_{L,\frac{N}{2}} &
        \mathbf{\bar{S}}_{kgu}^{ hh}
    \end{bmatrix}^H  
    \begin{bmatrix}
        \bm{\Phi}_{kgu}^{vv} & \bm{\Phi}_{kgu}^{hv} \\
        \bm{\Phi}_{kgu}^{vh} & \bm{\Phi}_{kgu}^{hh}
    \end{bmatrix} 
    \setlength{\arraycolsep}{2pt} 
    \begin{bmatrix}
        \mathbf{\bar{G}}_{kgu}^{ v v} &
        \sqrt{\chi^{\text{\tiny BS-IRS}}}
        \mathbf{\bar{G}}_{kgu}^{ h v} \\
        \sqrt{\chi^{\text{\tiny BS-IRS}}}
        \mathbf{\bar{G}}_{kgu}^{ v h} &
        \mathbf{\bar{G}}_{kgu}^{ h h}
    \end{bmatrix} 
    \nonumber\\[-1.5mm]
    &+
    \sqrt{\zeta_{kgu}^\text{\tiny BS-U}}\renewcommand*{\arraystretch}{1}
    \begin{bmatrix}
        \mathbf{\bar{D}}_{kgu}^{ v v} &
        \sqrt{\chi^{\text{\tiny BS-U}}}
        \mathbf{\bar{D}}_{kgu}^{ v h} \\
        \sqrt{\chi^{\text{\tiny BS-U}}}
        \mathbf{\bar{D}}_{kgu}^{ h v} &
        \mathbf{\bar{D}}_{kgu}^{ h h}
    \end{bmatrix}^H  \in \mathbb{C}^{N\times M},
\end{align}
where $\mathbf{\bar{D}}_{kgu}^{ pq} \in \mathbb{C}^{\frac{M}{2} \times \frac{N}{2}}$, $\mathbf{\bar{S}}_{kgu}^{ pq} \in \mathbb{C}^{L\times \frac{N}{2}}$, and $\mathbf{\bar{G}}_{kgu}^{ pq} \in \mathbb{C}^{L\times \frac{M}{2}}$ model, respectively, the fast-fading channels between the BS and the $u$th user (link BS-U), the $u$th IRS and the $u$th user (link IRS-U), and the BS and the $u$th IRS (link BS-IRS), from the polarization $p$ to the polarization $q$, in which $p,q\in \{v,h\}$, with $\chi^{\text{\tiny BS-U}}$ and $\chi^{\text{\tiny BS-IRS}} \in[0,1]$ denoting the inverse of the cross-polar discrimination parameter (iXPD) that measures the power leakage between polarizations in the links BS-U and BS-IRS. Moreover, $\frac{1}{\sqrt{2}}$ is the energy normalization factor, and $\zeta_{kgu}^\text{\tiny BS-U}$, $\zeta_{kgu}^\text{\tiny IRS-U}$, and $\zeta_{kgu}^\text{\tiny BS-IRS}$ represents the large-scale fading coefficients for the links BS-U, IRS-U, and BS-IRS, respectively. Observe that, the channel in \eqref{eq:ch0} consists of a generalization of that introduced in Section \ref{secfund}, with the difference that now both the transmitter, i.e., the BS, and receivers employ multiple dual-polarized antennas. Also, notice that we model depolarization phenomena in the links BS-U and BS-IRS, but not in the link IRS-U\footnote{Although depolarization phenomena are not considered in the link IRS-U, we would like to emphasize that the proposed model can be easily extended to this more general case. However, such consideration would lead to a more intricate mathematical formulation of difficult interpretation. Therefore, we choose not to address this issue in this work.}. This means that only negligible power leaks between polarizations in the propagation channels between the IRSs and users.

Furthermore, due to the closely spaced antennas at the BS and due to the scattering environment surrounding each spatial cluster, we assume that $\mathbf{\bar{D}}_{kgu}^{ pq}$, and $\mathbf{\bar{G}}_{kgu}^{ pq}$ are correlated, i.e., are rank deficient. On the other hand, we model $\mathbf{\bar{S}}_{kgu}^{ pq}$ as a full rank channel matrix. Under such assumptions, the covariance matrices of the links BS-IRS and BS-U can be calculated as \cite{ni3}
\begin{align}
&\mathbf{R}^{\text{\tiny BS-IRS}}_{k} = \zeta_{kgu}^\text{\tiny BS-IRS}(\chi^{\text{\tiny BS-IRS}}+1) \mathbf{I}_2 \otimes \mathbf{R}_k, \label{covmat1}\\[-2mm]
&\mathbf{R}^{\text{\tiny BS-U}}_{k} = \zeta_{kgu}^\text{\tiny BS-U}(\chi^{\text{\tiny BS-U}}+1)
\mathbf{I}_2 \otimes \mathbf{R}_k,\label{covmat2}
\end{align}
where $\mathbf{R}_k$ is the covariance matrix observed in each polarization, with rank denoted by $r_k$. Note that, we have assumed that the links BS-U and BS-IRS share the same covariance matrix $\mathbf{R}_k$. This is valid for the scenario where both IRS and users are located within the same cluster of scatterers, which in our model is reasonable since the user is located nearby its serving IRS.

Recalling the Karhunen–Loève representation \cite{ref1}, the channel in \eqref{eq:ch0} can be rewritten as
\begin{align}\label{eq:ch1} 
    \mathbf{H}^H_{kgu} &= 
    \left(\setlength{\arraycolsep}{3pt} \renewcommand*{\arraystretch}{1}
    \begin{bmatrix}
        \mathbf{S}_{kgu}^{ vv} &
        \mathbf{0}_{L,\frac{N}{2}} \\
        \mathbf{0}_{L,\frac{N}{2}} &
        \mathbf{S}_{kgu}^{ hh}
    \end{bmatrix}^H \hspace{-1mm}    
    \begin{bmatrix}
        \bm{\Phi}_{kgu}^{vv} & \bm{\Phi}_{kgu}^{hv} \\
        \bm{\Phi}_{kgu}^{vh} & \bm{\Phi}_{kgu}^{hh}
    \end{bmatrix} 
    \setlength{\arraycolsep}{2pt}
    \begin{bmatrix}
        \mathbf{G}_{kgu}^{ v v} &
        \mathbf{G}_{kgu}^{ h v} \\
        \mathbf{G}_{kgu}^{ v h} &
        \mathbf{G}_{kgu}^{ h h}
    \end{bmatrix} + 
    \begin{bmatrix}
        \mathbf{D}_{kgu}^{ v v} &
        \mathbf{D}_{kgu}^{ v h} \\
        \mathbf{D}_{kgu}^{ h v} &
        \mathbf{D}_{kgu}^{ h h}
    \end{bmatrix}^H \right) \left(\mathbf{I}_2 \kron \left( \mathbf{\Lambda}_{k}^{\frac{1}{2}} \mathbf{U}^H_{k} \right) \right)  \nonumber\\[-1mm]
    & =  \left(\mathbf{S}_{kgu}^H
      \bm{\Theta}_{kgu}
    \mathbf{G}_{kgu} + \mathbf{D}_{kgu}^H \right)  \left(\mathbf{I}_2 \kron \left( \mathbf{\Lambda}_{k}^{\frac{1}{2}} \mathbf{U}^H_{k} \right) \right),
\end{align}
where $\mathbf{\Lambda}_{k} \in \mathbb{R}^{r_k^\star \times r_k^\star}_{>0}$ is a diagonal matrix that collects $r_k^\star$ nonzero eigenvalues of $\mathbf{R}_k$, sorted in descending order, $\mathbf{U}_{k} \in \mathbb{C}^{\frac{M}{2} \times r_k^\star}$ is a unitary matrix containing the first $r_k^\star$ left eigenvectors of $\mathbf{R}_k$, corresponding to the eigenvalues in $\mathbf{\Lambda}_{k}$, $\mathbf{S}_{kgu}^{ pq} \in \mathbb{C}^{L\times\frac{N}{2}}$ is the full rank channel matrix of the link IRS-U, and $\mathbf{D}_{kgu}^{ pq} \in \mathbb{C}^{r_k^\star \times \frac{N}{2}}$ and $\mathbf{G}_{kgu}^{ pq} \in \mathbb{C}^{L\times r_k^\star }$ represent, respectively, the reduced-dimension fast-fading channels of the links BS-U and BS-IRS, from the polarization $p$ to the polarization $q$, with $p,q\in \{v,h\}$, whose entries follow the complex Gaussian distribution with zero mean and unit variance. Note that, for notation simplicity, the iXPD, the large scale fading coefficients, and the normalization factor $\frac{1}{\sqrt{2}}$ have been absorbed in the corresponding channel matrices.

With the above channel model, after the superimposed symbols have propagated through all wireless links, the $u$th user in the $g$th group within the $k$th cluster observes the following signal
\begin{align}\label{eq04}
\mathbf{y}_{kgu} &= \left(\mathbf{S}_{kgu}^H
    \hspace{-1mm}   \bm{\Theta}_{kgu}
    \mathbf{G}_{kgu} \hspace{-1mm} + \hspace{-1mm} \mathbf{D}_{kgu}^H \right) \hspace{-1.3mm} \left(\mathbf{I}_2 \kron \left( \mathbf{\Lambda}_{k}^{\frac{1}{2}} \mathbf{U}^H_{k} \right) \right)
 \hspace{-1.5mm} \sum_{m=1}^{K} \mathbf{P}_m \hspace{-1.3mm}\sum_{n=1}^{G} \sum_{i=1}^{U} \mathbf{v}_{mni} \alpha_{mni}{x}_{mni} + \renewcommand*{\arraystretch}{1}\begin{bmatrix} \mathbf{n}^v_{kgu} \\ \mathbf{n}^h_{kgu} \end{bmatrix}, %\in \mathbb{C}^{N \times 1},
\end{align}
where $\mathbf{n}^p_{kgu} \in \mathbb{C}^{\frac{N}{2}\times 1}$ is the noise vector observed at the receive antennas of polarization $p\in \{v,h\}$, whose entries follow the complex Gaussian distribution with zero mean and variance $\sigma_n$. 

Next, we provide details on the design of the precoding matrices, IRS optimization, and detection strategy.

\section{Precoding, IRS Optimization, and reception matrices}

\subsection{Spatial interference cancellation}\label{ssec:prec}
As mentioned before, the precoding matrix $\mathbf{P}_k$ is intended to remove the interference of different spatial clusters. From the signal model in \eqref{eq04}, it is clear that this objective can be accomplished if $\left[\mathbf{I}_2 \otimes \left( \mathbf{\Lambda}_{k}^{\frac{1}{2}} \mathbf{U}^H_{k} \right) \right]  \mathbf{P}_k = \mathbf{0}, \forall k'\neq k$, i.e., $\mathbf{P}_k$ should be orthogonal to the subspace spanned by the left eigenvectors of interfering clusters. Therefore, $\mathbf{P}_k$ can be computed from the null space of the matrix $\bm{\Omega}_k = [\mathbf{U}_{1}, \cdots, \mathbf{U}_{k-1}, \mathbf{U}_{k+1}, \cdots, \mathbf{U}_{K}] \in \mathbb{C}^{\frac{M}{2} \times \sum_{k'\neq k} r^\star_{k'}}$. This task can be performed by exploiting the singular value decomposition (SVD) of $\bm{\Omega}_k$. Specifically, the left eigenvectors of $\bm{\Omega}_k$ obtained from its SVD can be partitioned as $\mathbf{\Tilde{U}}_k = \begin{bmatrix} \mathbf{\Tilde{U}}^{(1)}_k & \mathbf{\Tilde{U}}^{(0)}_k \end{bmatrix}$, with $\mathbf{\Tilde{U}}^{(0)}_k \in \mathbb{C}^{\frac{M}{2}\times \frac{M}{2} - \sum_{k'\neq k} r^\star_{k'}}$ being a unitary matrix composed by the left eigenvectors of $\bm{\Omega}_k$ associated with its last $\frac{M}{2} - \sum_{k'\neq k} r^\star_{k'}$ vanishing eigenvalues. Since the columns of $\mathbf{\Tilde{U}}^{(0)}_k$ form a set of orthonormal basis vectors for the null space of $\bm{\Omega}_k$, we have that $\mathbf{H}^H_{k'gu}(\mathbf{I}_2 \otimes \mathbf{\Tilde{U}}^{(0)}_k) = 0$, $\forall k' \neq k$. Therefore, the goal of nulling out inter-cluster interference can be already fulfilled by constructing $\mathbf{P}_k$ from the columns of $\mathbf{\Tilde{U}}^{(0)}_k$. However, following the strategy proposed in \cite{ref6}, and given \eqref{eq04}, we can further improve the performance of the system by matching $\mathbf{P}_k$ to the dominant eigenmodes of the matrix $\mathbf{\Pi}_k = \mathbf{I}_2 \otimes  \left[\left( \mathbf{\Tilde{U}}^{(0)}_k \right)^H  \left(\mathbf{U}_{k}  \mathbf{\Lambda}_{k}^{\frac{1}{2}} \right) \right]$. This can be accomplished by multiplying $\mathbf{\Tilde{U}}^{(0)}_k$ by a unitary matrix constructed from the dominant eigenvectors of the covariance matrix of $\mathbf{\Pi}_k$, i.e.,  from $\mathbf{\Pi}_k (\mathbf{\Pi}_k)^H  = \mathbf{I}_2 \otimes \left[\left( \mathbf{\Tilde{U}}^{(0)}_k \right)^H \mathbf{R}_k \mathbf{\Tilde{U}}^{(0)}_k \right] = \mathbf{I}_2 \otimes \mathbf{\tilde{\Xi}}_{k}$. To be more specific, by representing the left eigenvectors of $\mathbf{\tilde{\Xi}}_{k}$ by $\mathbf{\bar{U}}_k = \begin{bmatrix} \mathbf{\bar{U}}^{(1)}_k & \mathbf{\bar{U}}^{(0)}_k \end{bmatrix}$, with $\mathbf{\bar{U}}^{(1)}_k \in \mathbb{C}^{\left( \frac{M}{2} - \sum_{k'\neq k} r^\star_{k'} \right)\times \frac{\bar{M}}{2} } $ collecting the first $\frac{\bar{M}}{2}$ columns of $\mathbf{\bar{U}}_k$, the desired precoding matrix can be finally computed by $\mathbf{P}_k = \mathbf{I}_2 \otimes \left( \mathbf{\Tilde{U}}^{(0)}_k \mathbf{\bar{U}}^{(1)}_k \right) = \mathbf{I}_2 \otimes \mathbf{\Tilde{P}}_k \in \mathbb{C}^{ M \times \bar{M}}$, in which, due to the dimensions of $\mathbf{\tilde{U}}_k$ and $\mathbf{\bar{U}}_k$, the constraints $K \leq \bar{M} \leq \left( M - 2\sum_{k'\neq k} r^\star_{k'} \right)$ and $\bar{M} \leq 2r^\star_{k}$ must be satisfied. \vspace{-2mm}

\subsection{Polarization assignment and formation of subsets} \label{passi}
In this subsection, we provide details on the strategy adopted for the formation of the polarization subsets and on the construction of the inner precoding vector $\mathbf{v}_{kgu}$. First, the BS sorts the users within each group in ascending order based on their large-scale fading coefficients observed in the link BS-U, such that $\zeta^{\text{\tiny BS-U}}_{kg1} < \zeta^{\text{\tiny BS-U}}_{kg2} < \cdots < \zeta^{\text{\tiny BS-U}}_{kgU}$. Then, without loss of generality, by assuming that $U$ is an even number, and aiming to form subsets with relatively balanced performance, users associated with odd indexes are assigned to the vertical polarization, and users associated with even indexes to the horizontal polarization, resulting in two disjoint subsets, the vertical subset $\mathcal{U}^{v} = \{1, 3,\cdots, U - 1\}$, containing $U^v = U/2$ users, and the horizontal subset $\mathcal{U}^{h} = \{2, 4, \cdots, U\}$, containing $U^h = U - U^v = U/2$ users. As a result, users within vertical subsets will be sorted as $\zeta^{\text{\tiny BS-U}}_{kg1} < \zeta^{\text{\tiny BS-U}}_{kg3} < \cdots < \zeta^{\text{\tiny BS-U}}_{kg(U-1)}$, and the ones within horizontal subsets as $\zeta^{\text{\tiny BS-U}}_{kg2} < \zeta^{\text{\tiny BS-U}}_{kg4} < \cdots < \zeta^{\text{\tiny BS-U}}_{kgU}$. In order to implement this strategy, for $1 \leq g \leq G$ and $1 \leq u \leq U$, the BS employs the following precoding vector
\begin{align}\label{pvec}
    \mathbf{v}_{kgu} = \renewcommand*{\arraystretch}{1}\begin{bmatrix}
    \mathbf{v}^v_{kgu} \\
    \mathbf{v}^h_{kgu}
    \end{bmatrix} =
    \renewcommand*{\arraystretch}{1.2}
    \begin{bmatrix} 
    \left[\mathbf{0}_{1,g-1}, \mathbf{1}_{\mathcal{U}^{v}}(u), \mathbf{0}_{1,\frac{\bar{M}}{2}-g}\right]^T \\
    \left[\mathbf{0}_{1,g-1}, \mathbf{1}_{\mathcal{U}^{h}}(u), \mathbf{0}_{1,\frac{\bar{M}}{2}-g}\right]^T
    \end{bmatrix},
\end{align}
where $\mathbf{1}_{\mathcal{A}}(i)$ is the indicator function of a subset $\mathcal{A}$, which results $1$ if $i \in \mathcal{A}$, and $0$ if $i \notin \mathcal{A}$. Note that, due to the structure of $\mathbf{v}_{kgu}$, the constraint $G\leq \bar{M}/2$ must be satisfied.

%We would like to highlight that
More sophisticated strategies for creating polarization subsets can be easily employed with the above precoding choice. The topic of user grouping in NOMA has been widely studied in the literature~\cite{wChen2020,dtDO2020}. However, our objective is not to develop an optimal user grouping strategy, but to shed light on the fundamental performance gains that our proposed scheme can render. Therefore, other possibilities go beyond the goals of this work. \vspace{-3mm}

\subsection{IRS optimization}
With the precoding matrix designed in the Section \ref{ssec:prec}, all inter-cluster interference can be effectively eliminated. Therefore, from now on, by focusing on the first cluster, we can drop the cluster subscript and simplify the signal in \eqref{eq04} as 
\begin{align}\label{sigtop}
    \mathbf{y}_{gu} &=\renewcommand*{\arraystretch}{.8} \left(\begin{bmatrix}
            \left[ (\mathbf{S}_{gu}^{ vv})^H
        \bm{\Phi}_{gu}^{vv}
        \mathbf{G}_{gu}^{ vv}
        +
        (\mathbf{S}_{gu}^{ vv})^H
        \bm{\Phi}_{gu}^{hv}
        \mathbf{G}_{gu}^{ vh} \right] & \left[ (\mathbf{S}_{gu}^{ vv})^H
        \bm{\Phi}_{gu}^{vv}
        \mathbf{G}_{gu}^{ hv}
        +
        (\mathbf{S}_{gu}^{ vv})^H
        \bm{\Phi}_{gu}^{hv}
        \mathbf{G}_{gu}^{ hh} \right] \\
         \left[ (\mathbf{S}_{gu}^{ hh})^H
        \bm{\Phi}_{gu}^{vh}
        \mathbf{G}_{gu}^{ vv}
        +
        (\mathbf{S}_{gu}^{ hh})^H
        \bm{\Phi}_{gu}^{hh}
        \mathbf{G}_{gu}^{ vh} \right] & \left[ (\mathbf{S}_{gu}^{ hh})^H
        \bm{\Phi}_{gu}^{vh}
        \mathbf{G}_{gu}^{ hv}
        +
        (\mathbf{S}_{gu}^{ hh})^H
        \bm{\Phi}_{gu}^{hh}
        \mathbf{G}_{gu}^{ hh} \right]
    \end{bmatrix}\right.\nonumber\\[-1mm]
    &+\left. \renewcommand*{\arraystretch}{1}
\begin{bmatrix}
        (\mathbf{D}_{gu}^{ v v})^H &
        (\mathbf{D}_{gu}^{ h v})^H \\
        (\mathbf{D}_{gu}^{ v h})^H &
        (\mathbf{D}_{gu}^{ h h})^H
    \end{bmatrix}\right) \renewcommand*{\arraystretch}{1}
    \begin{bmatrix}
        \mathbf{\Lambda}^{\frac{1}{2}} \mathbf{U}^H \mathbf{\Tilde{P}} & \mathbf{0}_{\frac{M}{2}, \frac{\bar{M}}{2}} \\
        \mathbf{0}_{\frac{M}{2}, \frac{\bar{M}}{2}} &
        \mathbf{\Lambda}^{\frac{1}{2}} \mathbf{U}^H \mathbf{\Tilde{P}}
    \end{bmatrix}
  \sum_{n=1}^{G} \sum_{i=1}^{U} 
  \begin{bmatrix}
    \mathbf{v}^v_{ni} \\
    \mathbf{v}^h_{ni}
    \end{bmatrix}
  \alpha_{ni}{x}_{ni} + \begin{bmatrix} \mathbf{n}^v_{gu} \\ \mathbf{n}^h_{gu} \end{bmatrix}.
\end{align}

As can be observed in \eqref{sigtop}, in both the BS-U and the BS-IRS-U links, the symbols intended to the subsets assigned to the vertical polarization propagate through the channels modeled by the left blocks of the channel matrices, while the symbols for subsets assigned to the horizontal polarization propagate through the right blocks. Therefore, the IRSs of users assigned to the vertical polarization should be optimized to null out the right channel blocks, and the IRSs for users assigned to the horizontal polarization should null out the left channel blocks. More specifically, we aim to achieve in subsets assigned to the vertical polarization:
\begin{align}\label{objvert}
    &\renewcommand*{\arraystretch}{1}\begin{bmatrix}
       (\mathbf{S}_{gu}^{ vv})^H
        \bm{\Phi}_{gu}^{vv}
        \mathbf{G}_{gu}^{ hv}
        +
        (\mathbf{S}_{gu}^{ vv})^H
        \bm{\Phi}_{gu}^{hv}
        \mathbf{G}_{gu}^{ hh} \\
       (\mathbf{S}_{gu}^{ hh})^H
        \bm{\Phi}_{gu}^{vh}
        \mathbf{G}_{gu}^{ hv}
        +
        (\mathbf{S}_{gu}^{ hh})^H
        \bm{\Phi}_{gu}^{hh}
        \mathbf{G}_{gu}^{ hh}
    \end{bmatrix} +
    \begin{bmatrix}
        (\mathbf{D}_{gu}^{ h v})^H \\
        (\mathbf{D}_{gu}^{ h h})^H
    \end{bmatrix}
    \approx \begin{bmatrix}
        \mathbf{0}_{\frac{N}{2},r^\star_k} \\
        \mathbf{0}_{\frac{N}{2},r^\star_k}
    \end{bmatrix},
\end{align}
and in subsets assigned to the horizontal polarization:
\begin{align}
    &\renewcommand*{\arraystretch}{1}\begin{bmatrix}
        (\mathbf{S}_{gu}^{ vv})^H
        \bm{\Phi}_{gu}^{vv}
        \mathbf{G}_{gu}^{ vv}
        +
        (\mathbf{S}_{gu}^{ vv})^H
        \bm{\Phi}_{gu}^{hv}
        \mathbf{G}_{gu}^{ vh}
      \\
        (\mathbf{S}_{gu}^{ hh})^H
        \bm{\Phi}_{gu}^{vh}
        \mathbf{G}_{gu}^{ vv}
        +
        (\mathbf{S}_{gu}^{ hh})^H
        \bm{\Phi}_{gu}^{hh}
        \mathbf{G}_{gu}^{ vh}
    \end{bmatrix} +
    \begin{bmatrix}
        (\mathbf{D}_{gu}^{ v v})^H \\
        (\mathbf{D}_{gu}^{ v h})^H
    \end{bmatrix}
    \approx \begin{bmatrix}
        \mathbf{0}_{\frac{N}{2},r^\star_k} \\
        \mathbf{0}_{\frac{N}{2},r^\star_k}
    \end{bmatrix}.
\end{align}

Note that, by mitigating the transmissions originated from the interfering polarization, we can transform depolarization phenomena, which usually are harmful, into an advantage. More specifically, this strategy should enable users to receive their intended messages, transmitted from a single polarization (or vertical, or horizontal), in both receive polarizations, ideally, interference-free. Take a user within a vertical subset, for instance. If all interference from the horizontal subset can be canceled, the message transmitted from the vertical polarization at the BS will reach this user through both vertical-to-vertical co-polar transmissions and vertical-to-horizontal cross-polar transmissions. In other words, the proposed scheme enables polarization diversity, as anticipated in previous sections.

Given that the objectives for the IRSs of vertical and horizontal polarization subsets are similar, i.e., to null out co-polar and cross-polar transmissions from interfering subsets, the optimization procedure for both subsets will be also similar. For this reason, and also due to space constraints, we focus on the optimization of IRSs for subsets assigned for vertical polarization. Specifically, based on \eqref{objvert}, the reflecting coefficients for users assigned to the vertical polarization can be optimized by solving the following problem
\begin{subequations}\label{p1}
\begin{align}
        &\underset{\bm{\Phi}_{gu}^{hv}, \bm{\Phi}_{gu}^{hh} }{\underset{\bm{\Phi}_{gu}^{vv}, \bm{\Phi}_{gu}^{vh},}{\min}}
        \left\| \renewcommand*{\arraystretch}{1}
        \begin{bmatrix}
       (\mathbf{S}_{gu}^{ vv})^H
        \bm{\Phi}_{gu}^{vv}
        \mathbf{G}_{gu}^{ hv} \\
       (\mathbf{S}_{gu}^{ hh})^H
        \bm{\Phi}_{gu}^{vh}
        \mathbf{G}_{gu}^{ hv}
    \end{bmatrix}  + \begin{bmatrix}
        (\mathbf{S}_{gu}^{ vv})^H
        \bm{\Phi}_{gu}^{hv}
        \mathbf{G}_{gu}^{ hh} \\
        (\mathbf{S}_{gu}^{ hh})^H
        \bm{\Phi}_{gu}^{hh}
        \mathbf{G}_{gu}^{ hh}
    \end{bmatrix} +
    \begin{bmatrix}
        (\mathbf{D}_{gu}^{ h v})^H \\
        (\mathbf{D}_{gu}^{ h h})^H
    \end{bmatrix}
         \right\|^2 \\[-2mm]
    &\quad\text{s.t.} \hspace{2mm} |\omega^{pq}_{gu,l}|^2 \leq 1, \hspace{2mm}  \forall l \in [1, L], \forall p,q \in \{v,h\}, \label{p12b} \\[-2mm]
    &\quad \quad \bm{\Phi}_{gu}^{vv}, \bm{\Phi}_{gu}^{vh}, \bm{\Phi}_{gu}^{hv}, \bm{\Phi}_{gu}^{hh} \text{ diagonal.}
\end{align}
\end{subequations}
where \eqref{p12b} is the constraint for ensuring a passive reflection. The problem above can be seen as a generalization of the unconstrained least squares problem for matrix equations, in which some studies have been carried out in \cite{Shim03}. However, due to the element-wise quadratic constraint and the diagonal matrices constraint, it becomes difficult to solve \eqref{p1} in its current form. To overcome this challenge, next, we transform \eqref{p1} in an equivalent tractable problem.

Using the Khatri-Rao identity $(\mathbf{C}^T \odot \mathbf{A}) \textit{vecd}\{\mathbf{B}\} = \textit{vec}\{ \mathbf{A}\mathbf{B}\mathbf{C}\} $ \cite{Brewer78}, we define:
\begin{align*}
  &\bm{\theta}^{pq}_{gu} = \textit{vecd}\{\bm{\Phi}^{pq}_{gu}\} \in \mathbb{C}^{ L \times 1},
  \hspace{3mm}
  \mathbf{d}^{hv}_{gu} = \textit{vec}\left\{(\mathbf{D}_{gu}^{ h v})^H \right\} \in \mathbb{C}^{ \frac{N}{2} r_k^\star \times 1},
  \hspace{3mm}
  \mathbf{d}^{hh}_{gu} = \textit{vec}\left\{(\mathbf{D}_{gu}^{ h h})^H \right\} \in \mathbb{C}^{ \frac{N}{2} r_k^\star \times 1},
   \\[-2mm]
   &\mathbf{K}^{hv,vv}_{gu} = [(\mathbf{G}_{gu}^{ hv})^T \kr (\mathbf{S}_{gu}^{ vv})^H] \in \mathbb{C}^{ \frac{N}{2} r_k^\star \times L},
   \hspace{10mm}
   \mathbf{K}^{hh,vv}_{gu} = [(\mathbf{G}_{gu}^{ hh})^T \kr (\mathbf{S}_{gu}^{ vv})^H] \in \mathbb{C}^{ \frac{N}{2} r_k^\star \times L},
   \\[-2mm]
   &\mathbf{K}^{hv,hh}_{gu} = [(\mathbf{G}_{gu}^{ hv})^T \kr (\mathbf{S}_{gu}^{ hh})^H] \in \mathbb{C}^{ \frac{N}{2} r_k^\star \times L},
   \hspace{10mm}
   \mathbf{K}^{hh,hh}_{gu} = [(\mathbf{G}_{gu}^{ hh})^T \kr (\mathbf{S}_{gu}^{ hh})^H] \in \mathbb{C}^{ \frac{N}{2} r_k^\star \times L}.
\end{align*}
Then, we can transform \eqref{p1} into the following two sub-problems
\begin{subequations}\label{p2}
\begin{align}
        &\underset{\bm{\theta}^{vv}_{gu}, \bm{\theta}^{hv}_{gu}}{\min}
         \left\|\setlength{\arraycolsep}{1pt}
        \begin{bmatrix}
       \mathbf{K}^{hv,vv}_{gu} &
        \mathbf{K}^{hh,vv}_{gu} \end{bmatrix} \left[(\bm{\theta}^{vv}_{gu})^T, (\bm{\theta}^{hv}_{gu})^T\right]^T \hspace{-2mm}  +
    \mathbf{d}^{hv}_{gu}
        \right\|^2 \label{p2a}\\[-2mm]
    &\quad\text{s.t.} \hspace{2mm} \left\| \left[(\bm{\theta}^{vv}_{gu})^T, (\bm{\theta}^{hv}_{gu})^T \right]^T \right\|^2_{\infty} \leq 1,\label{p2b}
\end{align}
\end{subequations}\vspace{-3mm}
\begin{subequations}\label{p3}
\begin{align}
        &\underset{\bm{\theta}^{vh}_{gu}, \bm{\theta}^{hh}_{gu} }{\min}
         \left\| \setlength{\arraycolsep}{1pt}
        \begin{bmatrix}
       \mathbf{K}^{hv,hh}_{gu} &
       \mathbf{K}^{hh,hh}_{gu} \end{bmatrix} \left[(\bm{\theta}^{vh}_{gu})^T, (\bm{\theta}^{hh}_{gu})^T \right]^T  \hspace{-2mm}  +
    \mathbf{d}^{hh}_{gu} 
        \right\|^2 \label{p3a}\\[-2mm]
    &\quad\text{s.t.} \hspace{2mm} \left\| \left[(\bm{\theta}^{vh}_{gu})^T, (\bm{\theta}^{hh}_{gu})^T \right]^T \right\|^2_{\infty} \leq 1.\label{p3b}
\end{align}
\end{subequations}
which consist of least squares problems with $\mathpzc{L}_\infty$ norm constraints. Before we can solve the problems above, let us denote
$\mathbf{\bar{K}}_{gu} = \begin{bmatrix} \mathbf{K}^{hv,vv}_{gu} & \mathbf{K}^{hh,vv}_{gu} \end{bmatrix}$, $\mathbf{\bar{C}}_{gu} = \mathbf{\bar{K}}_{gu}^H\mathbf{\bar{K}}_{gu}$, and
$\mathbf{\Tilde{K}}_{gu} =  \begin{bmatrix} \mathbf{K}^{hv,hh}_{gu} & \mathbf{K}^{hh,hh}_{gu} \end{bmatrix}$, $\mathbf{\Tilde{C}}_{gu} = \mathbf{\Tilde{K}}_{gu}^H\mathbf{\Tilde{K}}_{gu}$, and rewrite the left-hand side of the constraints in \eqref{p2b} and \eqref{p3b}, respectively, as 
$\left\| \left[(\bm{\theta}^{vv}_{gu})^T \hspace{0mm} , (\bm{\theta}^{hv}_{gu})^T \right]^T \right\|^2_{\infty} \hspace{-1mm}=$ $\left[(\bm{\theta}^{vv}_{gu})^H \hspace{-1mm}, (\bm{\theta}^{hv}_{gu})^H \right] \hspace{0mm} \mathbf{B}_l \hspace{0mm} \left[(\bm{\theta}^{vv}_{gu})^T \hspace{-1mm}, (\bm{\theta}^{hv}_{gu})^T \right]^T$, and
$\left\| \left[(\bm{\theta}^{vh}_{gu})^T \hspace{-2mm} , (\bm{\theta}^{hh}_{gu})^T \right]^T \right\|^2_{\infty} \hspace{-3mm}=$ $\left[(\bm{\theta}^{vh}_{gu})^H \hspace{-1mm}, (\bm{\theta}^{hh}_{gu})^H \right] \hspace{-1mm}\mathbf{B}_l \hspace{-1mm}\left[(\bm{\theta}^{vh}_{gu})^T\hspace{-1mm}, (\bm{\theta}^{hh}_{gu})^T \right]^T \hspace{-1mm}$, where $\mathbf{B}_l = \textit{diag}\{\mathbf{e}_l\}, l=1,\cdots, L$, with $\mathbf{e}_l$ representing the standard basis vector that contains 1 in the $l$th position and zeros elsewhere.
Then, by expanding the objective functions in \eqref{p2a} and \eqref{p3a}, we obtain
\begin{subequations}\label{p4}
\begin{align}
        \underset{\bm{\theta}^{vv}_{gu}, \bm{\theta}^{hv}_{gu}}{\min}
         & \left\{ \renewcommand{\arraystretch}{0.9}
         \begin{bmatrix}
                \bm{\theta}^{vv}_{gu} \\
                \bm{\theta}^{hv}_{gu}
         \end{bmatrix}^H \hspace{-2mm}
         \mathbf{\bar{C}}_{gu} \begin{bmatrix}
                \bm{\theta}^{vv}_{gu} \\
                \bm{\theta}^{hv}_{gu}
         \end{bmatrix} + 2\Re\hspace{-.6mm}\left\{(\mathbf{d}^{hv}_{gu})^H\mathbf{\bar{K}}_{gu}\hspace{-1mm} \begin{bmatrix}
                \bm{\theta}^{vv}_{gu} \\
                \bm{\theta}^{hv}_{gu}
         \end{bmatrix} \right\} \hspace{-.6mm} + \hspace{-.6mm} (\mathbf{d}^{hv}_{gu})^H\mathbf{d}^{hv}_{gu}\right\}
         \\[-2mm]
    \quad\text{s.t.} &\hspace{1mm} \renewcommand{\arraystretch}{0.9} \begin{bmatrix}
                \bm{\theta}^{vv}_{gu} \\
                \bm{\theta}^{hv}_{gu}
         \end{bmatrix}^H\hspace{-2mm} \mathbf{B}_l \begin{bmatrix}
                \bm{\theta}^{vv}_{gu} \\
                \bm{\theta}^{hv}_{gu}
         \end{bmatrix} \leq 1,\label{p4b}
\end{align}
\end{subequations}\vspace{-1mm}
\begin{subequations}\label{p5}
\begin{align}
        \underset{\bm{\theta}^{vh}_{gu}, \bm{\theta}^{hh}_{gu}}{\min}&
         \left\{
         \renewcommand{\arraystretch}{0.9}
         \begin{bmatrix}
                \bm{\theta}^{vh}_{gu} \\
                \bm{\theta}^{hh}_{gu}
         \end{bmatrix}^H \hspace{-2mm}
         \mathbf{\Tilde{C}}_{gu} \begin{bmatrix}
                \bm{\theta}^{vh}_{gu} \\
                \bm{\theta}^{hh}_{gu}
         \end{bmatrix} + 2\Re\hspace{-.6mm}\left\{\hspace{-.6mm}(\mathbf{d}^{hh}_{gu})^H\mathbf{\Tilde{K}}_{gu}\hspace{-1mm} \begin{bmatrix}
                \bm{\theta}^{vh}_{gu} \\
                \bm{\theta}^{hh}_{gu}
         \end{bmatrix} \right\} \hspace{-.6mm} + \hspace{-.6mm} (\mathbf{d}^{hh}_{gu})^H\mathbf{d}^{hh}_{gu}\right\}
         \\[-2mm]
    \quad\text{s.t.}& \hspace{1mm} \renewcommand{\arraystretch}{0.9}
         \begin{bmatrix}
                \bm{\theta}^{vh}_{gu} \\
                \bm{\theta}^{hh}_{gu}
         \end{bmatrix}^H \hspace{-2mm} \mathbf{B}_l \begin{bmatrix}
                \bm{\theta}^{vh}_{gu} \\
                \bm{\theta}^{hh}_{gu}
         \end{bmatrix} \leq 1.\label{p5b}
\end{align}
\end{subequations}

It is straightforward to see that \eqref{p4} and \eqref{p5} are quadractically constrained quadratic problems. Given that the entries of $\mathbf{\bar{K}}_{gu}$ and $\mathbf{\Tilde{K}}_{gu}$ are independent complex Gaussian random variables, $\mathbf{\bar{C}}_{gu}$ and $\mathbf{\Tilde{C}}_{gu}$ will be positive semidefinite matrices with probability one. Furthermore, since $\mathbf{z}^H\mathbf{B}_l \mathbf{z} = |[\mathbf{z}]_{l}|^2 \geq 0, \hspace{0mm} \forall \mathbf{z} \in \mathbb{C}^{L\times 1}$, the matrix $\mathbf{B}_l$ is also positive semidefinite. As a result, the problems \eqref{p4} and \eqref{p5} are convex and, consequently, have global optimal solutions that can be efficiently computed via interior-points methods in polynomial time \cite{Luo2010}. Then, by denoting the optimal vectors of reflection coefficients by $\bm{\Dot{\theta}}^{pq}_{gu}$, obtained by solving \eqref{p4} and \eqref{p5}, the reflection matrices that minimizes \eqref{p1} are obtained as $\bm{\Phi}^{pq}_{gu} = \textit{diag}\left\{\bm{\Dot{\theta}}^{pq}_{gu}\right\}$, $\forall p,q \in \{v,h\}$.

Since the optimization problems in \eqref{p4} and \eqref{p5} depend on the fast fading channel matrices observed in all propagation links, one can wonder how the IRSs can be configured in a real-time manner. In fact, there are different approaches to perform such optimizations, which can require or not the knowledge of the channel stated information (CSI) on the IRSs, as explained in \cite{aswc2020}. For instance, if the installed IRSs have sensing capabilities, the channels in the reflected link BS-IRS-U can be estimated directly on them, and the optimization can run in the IRSs' local controllers in a distributed fashion. For this, the BS needs to inform the CSI of the direct link BS-U to the IRSs. The disadvantage of this strategy is that when the number of transmit/receive antennas and reflecting elements increases, the optimization becomes excessively complex for the limited processing power of the IRSs. As an alternative, it is possible to simplify the IRSs' hardware by removing the sensing components and transfer the burden of the channel estimations and the IRSs optimization entirely to the BS, which disposes of abundant computational resources. In this centralized approach, the channel matrices of the BS-IRS-U and BS-U links are both estimated in the BS, allowing the BS itself to compute the optimal sets of reflection coefficients. Then, after computing the reflecting coefficients, they are sent to the IRSs through an ultra-fast backhaul link. In particular, since we consider an IRS to be a nearly passive device with low computational capabilities, here we assume a centralized optimization. \vspace{-2mm}

\subsection{Signal reception}
Since we have already provided details on the optimization of the IRSs, for the sake of simplicity, hereinafter the links BS-IRS-U and BS-U are absorbed into a single channel matrix, and \eqref{eq:ch1} is rewritten in a more compact structure, as follows
\begin{align}\label{eq:ch24}
    \mathbf{H}^H_{gu} = \renewcommand*{\arraystretch}{1} \begin{bmatrix}
        \mathbf{\Tilde{H}}_{gu}^{v v} &
        \mathbf{\Tilde{H}}_{gu}^{v h} \\
        \mathbf{\Tilde{H}}_{gu}^{h v} &
        \mathbf{\Tilde{H}}_{gu}^{h h}
    \end{bmatrix}^H,
\end{align}
where $\mathbf{\Tilde{H}}_{gu}^{pq}$ accounts for both direct and reflected transmissions that depart the BS from polarization $p$ and arrive at the user's devices on polarization $q$, with $p,q \in \{v,h\}$, e.g., the effective vertical-to-vertical channel matrix is defined by $\mathbf{\Tilde{H}}_{gu}^{vv} =  \mathbf{U}_{k}  \mathbf{\Lambda}_{k}^{\frac{1}{2}} \left[ (\mathbf{S}_{gu}^{ vv})^H \bm{\Phi}_{gu}^{vv} \mathbf{G}_{gu}^{ vv}\right.$ + $\left.(\mathbf{S}_{gu}^{ vv})^H \bm{\Phi}_{gu}^{hv} \mathbf{G}_{gu}^{ vh} \right]^H$ $ + \hspace{1mm} \mathbf{U}_{k}  \mathbf{\Lambda}_{k}^{\frac{1}{2}} \mathbf{D}_{gu}^{ vv} $. With this notation, the signal in \eqref{sigtop} can be simplified to 
\begin{align}\label{sigrec1}
\mathbf{y}_{gu} =  \renewcommand*{\arraystretch}{1}
  \begin{bmatrix}
        (\mathbf{\Tilde{H}}_{gu}^{v v})^H\mathbf{\Tilde{P}} &
        (\mathbf{\Tilde{H}}_{gu}^{h v})^H\mathbf{\Tilde{P}} \\
        (\mathbf{\Tilde{H}}_{gu}^{v h})^H \mathbf{\Tilde{P}} &
        (\mathbf{\Tilde{H}}_{gu}^{h h})^H \mathbf{\Tilde{P}}
    \end{bmatrix}  \sum_{n=1}^{G}  \sum_{i=1}^{U} 
  \begin{bmatrix}
    \mathbf{v}^v_{ni} \\
    \mathbf{v}^h_{ni}
    \end{bmatrix}
  \alpha_{ni}{x}_{ni}
    + \begin{bmatrix} \mathbf{n}^v_{gu} \\ \mathbf{n}^h_{gu} \end{bmatrix} .
\end{align}

Then, in order to explain our detection strategy, without loss of generality, we focus on subsets assigned to the vertical polarization. Remember that the IRSs of users assigned to the vertical polarization are optimized to mitigate all transmissions originated at the BS from the horizontal polarization. Therefore, by relying on the effectiveness of the IRS, we exploit the left blocks of the channel matrix in \eqref{sigrec1} to construct our detection matrix. More specifically, in order to remove the remaining interference from other subsets also assigned to the vertical polarization, the $u$th user exploits the virtual channels $\mathbf{\underline{H}}_{gu}^{v v} = (\mathbf{\Tilde{H}}_{gu}^{v v})^H\mathbf{\Tilde{P}}$ and $\mathbf{\underline{H}}_{gu}^{v h} = (\mathbf{\Tilde{H}}_{gu}^{v h})^H\mathbf{\Tilde{P}}$ to construct the following detection matrix
\begin{align}\label{detecm}
    \mathbf{H}^{\dagger}_{gu} &= \renewcommand*{\arraystretch}{1} \begin{bmatrix}
           \mathbf{H}^{\dagger v}_{gu} & \mathbf{0}_{\frac{\bar{M}}{2}, \frac{N}{2}} \\
           \mathbf{0}_{\frac{\bar{M}}{2}, \frac{N}{2}} & \mathbf{H}^{\dagger h}_{gu}
    \end{bmatrix} = \setlength{\arraycolsep}{0pt}\begin{bmatrix}
           [(\mathbf{\underline{H}}_{gu}^{v v})^H \mathbf{\underline{H}}_{gu}^{v v}]^{-1}(\mathbf{\underline{H}}_{gu}^{v v})^H & \mathbf{0}_{\frac{\bar{M}}{2}, \frac{N}{2}} \\
           \mathbf{0}_{\frac{\bar{M}}{2}, \frac{N}{2}} & [(\mathbf{\underline{H}}_{gu}^{v h})^H \mathbf{\underline{H}}_{gu}^{v h}]^{-1}(\mathbf{\underline{H}}_{gu}^{v h})^H
    \end{bmatrix},
\end{align}
where $\mathbf{H}^{\dagger p}_{gu}$ is a left Moore–Penrose inverse intended to detect the signals impinging on the receive antennas with polarization $p$, in which it is assumed that $N\geq \bar{M}$. Then, after multipliying the signal in \eqref{sigrec1} by $\mathbf{H}^{\dagger}_{gu}$, the $u$th user obtains the following data vector
\begin{align}\label{sigrec2}
\mathbf{\hat{x}}_{gu} &=\renewcommand*{\arraystretch}{1} \begin{bmatrix} \mathbf{\hat{x}}^{v}_{gu} \\ \mathbf{\hat{x}}^{h}_{gu} \end{bmatrix} =  \setlength{\arraycolsep}{3pt}
  \begin{bmatrix}
        \mathbf{x}^v +
        \mathbf{H}^{\dagger v}_{gu}\mathbf{\underline{H}}_{gu}^{h v} \mathbf{x}^h \\
        \mathbf{x}^v +
        \mathbf{H}^{\dagger h}_{gu}\mathbf{\underline{H}}_{gu}^{h h} \mathbf{x}^h
    \end{bmatrix} + \begin{bmatrix} \mathbf{H}^{\dagger v}_{gu}\mathbf{n}^v_{gu} \\ \mathbf{H}^{\dagger h}_{gu} \mathbf{n}^h_{gu} \end{bmatrix},
\end{align}
where, due to the precoding vector in \eqref{pvec}, $\mathbf{x}^v$ is given by
\begin{align}\label{vecv}
\mathbf{x}^v &= \renewcommand*{\arraystretch}{.5} \begin{bmatrix} 
            \sum_{i\in \mathcal{U}^{v}} \alpha_{1i}x_{1i} \vspace{-1mm} \\
            \vdots \vspace{-1mm} \\
            \sum_{i\in \mathcal{U}^{v}} \alpha_{Gi}x_{Gi}
         \end{bmatrix}.
\end{align}

Note in \eqref{sigrec2} that, by employing $\mathbf{H}^{\dagger}_{gu}$, users will obtain in both receive polarizations corrupted replicas of the vector of superimposed symbols that was transmitted by the BS from the vertical polarization. Moreover, as one can observe in \eqref{vecv}, each element of $\mathbf{x}^v$ consists of a superimposed symbol intended to a specific user subset. Therefore, a user within the $g$th vertical subset is able to decode its symbol from the $g$th element of both $\mathbf{\hat{x}}^{v}_{gu}$ and $\mathbf{\hat{x}}^{h}_{gu}$. In particular, inspired by the strategy proposed in \cite{ni3}, the data symbols will be decoded from the polarization that renders the highest effective channel gain, denoted in this work as the polarization $\ddot{p}$. As a result, the superimposed symbol recovered by the $u$th user in the $g$th vertical subset before carrying out SIC is given by
\begin{align}\label{datavec1}
    [\mathbf{\hat{x}}^{\ddot{p}}_{gu}]_{g} = \sum_{i\in \mathcal{U}^{v}} \alpha_{gi}x_{gi} +
        [\mathbf{H}^{\dagger \ddot{p}}_{gu}\mathbf{\underline{H}}_{gu}^{h \ddot{p}}\mathbf{x}^h]_{g} + [\mathbf{H}^{\dagger \ddot{p}}_{gu}\mathbf{n}^{\ddot{p}}_{gu}]_{g}.
\end{align}

Users within horizontal subsets employ the same strategy. However, differently from vertical subsets, the matrix $\mathbf{H}^{\dagger}_{gu}$ is constructed based on the right blocks of the channel matrix in \eqref{sigrec1}. \vspace{-2mm}

\section{Performance Analysis}
In this section, we carry out an in-depth study of the performance of the proposed system. By taking into account polarization interference and errors from imperfect SIC, we first provide a general expression for the SINR observed by each user during the SIC process. A statistical analysis is then performed to identify the distribution of the channel gains, which turns out to be challenging to find for general values of reflecting elements. We then investigate the limiting case for $L\rightarrow \infty$, in which the asymptotic distribution is determined. Lastly, by considering large values of $L$, a closed-form analytical expression for the ergodic rates is derived. \vspace{-2mm}

\subsection{SINR analysis}
Before the users can read their messages, they still need to decode the superimposed symbol in \eqref{datavec1} through SIC. Recall that due to the polarization assignment strategy proposed in the Section \ref{passi}, users within each subset are sorted in ascending order based on their large scale coefficients, e.g., in vertical subsets $\zeta^{\text{\tiny BS-U}}_{kg1} < \zeta^{\text{\tiny BS-U}}_{kg3} < \cdots < \zeta^{\text{\tiny BS-U}}_{kgU^v}$. As a result, following the NOMA protocol, before the $u$th user in the polarization subset $\mathcal{U}^{p}$, $p\in \{v,h\}$, can retrieve its own message, it carries out SIC to decode the symbol intended for the $m$th weaker user,  $\forall m < u,$ $m\in \mathcal{U}^{p}$, and treats the message to the $n$th stronger user as interference, $\forall n > u, n \in \mathcal{U}^{p}$. Ideally, the symbols intended for weaker users can be perfectly removed by SIC. However, as clarified in Section I, due to many factors, SIC errors are inevitable in practice. Therefore, users suffer from SIC error propagation in the proposed system, and this is modeled as a linear function of the power of decoded symbols, as in \cite{SenaISIC2020}. Then, after all SIC decodings, the $u$th user assigned to the polarization subset $\mathcal{U}^{p}$ in the $g$th group observes the following symbol
\begin{align}\label{sigbsic4}
    {\hat{x}}_{gu} &=\hspace{-1mm} \underset{\text{Desired symbol}}{\underbrace{ 
    \alpha_{gu}x_{gu}
    }} 
    + \hspace{-2mm}\underset{\text{Interference of stronger users}}{\underbrace{
    \sum_{m \in \{a | \hspace{.5mm} a > u, \hspace{.5mm} a\in \mathcal{U}^{p} \}} \hspace{-8mm} \alpha_{gm}x_{gm}
    }}
    + \underset{\text{Residual SIC interference}}{\underbrace{
    \sqrt{\xi} \hspace{-8mm} \sum_{n \in \{b | \hspace{.5mm} b < u, \hspace{.5mm} b\in \mathcal{U}^{p} \}} \hspace{-8mm} \alpha_{gn}x_{gn}
    }} 
   \hspace{-1mm} + \underset{\text{Polarization interference}}{\underbrace{
    [\mathbf{H}^{\dagger \ddot{p}}_{gu}\mathbf{\underline{H}}_{gu}^{t \ddot{p}} \mathbf{x}^t]_{g}
    }}
    \hspace{-2mm} + \underset{\text{Noise}}{\underbrace{
    [\mathbf{H}^{\dagger \ddot{p}}_{gu}\mathbf{n}^{\ddot{p}}_{gu}]_{g}
    }},
\end{align}
where the superscript $t$ represents the interfering polarization that is defined by $t= h$, if $u\in \mathcal{U}^{v}$, or $t = v$, if $u\in \mathcal{U}^{h}$, and $\xi \in [0,1]$ is the SIC error propagation factor, in which $\xi = 0$ corresponds to the perfect SIC case, and $\xi = 1$ represents the scenario of maximum error. Moreover, note that if the IRS of the $u$th can completely eliminate the transmissions coming from the horizontally polarized BS antennas, the polarization interference term in \eqref{sigbsic4} will disappear.

The SINR observed during each SIC decoding is defined in the following lemma.

\paragraph*{Lemma I} Under the assumption of imperfect SIC, the $u$th user in the $g$th group decodes the data symbol intended to the $i$th user, $\forall i \leq u$, $i\in \mathcal{U}^{p}$, with the following SINR
\begin{align}\label{sinreq}
\gamma_{gu}^{i} &= \frac{\rho\mathcal{\ddot{h}}_{gu} \alpha_{gi}^2}{\rho\mathcal{\ddot{h}}_{gu}\mathfrak{I}_{gi} + \rho\mathcal{\ddot{h}}_{gu}\mathfrak{X}_{gu} + 
1},
\end{align}
where $\mathcal{\ddot{h}}_{gu} = \max \{\mathcal{h}^{v}_{gu}, \mathcal{h}^{h}_{gu}\}$, with $\mathcal{h}^{p}_{gu} = [1/\mathbf{H}^{\dagger p}_{gu}(\mathbf{H}^{\dagger p}_{gu})^H]_{gg}$ being the effective channel observed in the polarization $p$, $\mathfrak{X}_{gu} = \left|[\mathbf{H}^{\dagger \ddot{p}}_{gu}\mathbf{\underline{H}}_{gu}^{t \ddot{p}}\mathbf{x}^t]_{g} \right|^2$ represents the residual polarization interference left by the IRS, in which, if $u\in \mathcal{U}^{v}$, $t = h$, and if $u\in \mathcal{U}^{h}$, $t=v$. The symbol $\rho = 1/\sigma_n^{2}$ represents the SNR, and $\mathfrak{I}_{gi}$ is the total SIC interference given by
\begin{align}
    \mathfrak{I}_{gi} &=
    \begin{cases}
        \sum_{m=i+1}^{\max\{\mathcal{U}^{p}\}} \alpha_{gm}^2, \hspace{-2mm} & \text{ if } i = \min\{\mathcal{U}^{p}\},\\
        \sum_{m=i+1}^{\max\{\mathcal{U}^{p}\}} \alpha_{gm}^2 + \xi \sum_{n=\min\{\mathcal{U}^{p}\}}^{i-1} \alpha_{gn}^2, \hspace{-2mm}& \text{ if } \min\{\mathcal{U}^{p}\} < i \leq u < \max\{\mathcal{U}^{p}\},\\
        \xi \sum_{n=1}^{i-1} \alpha_{gn}^2, \hspace{-2mm} & \text{ if } i = u = \max\{\mathcal{U}^{p}\},
    \end{cases}
\end{align}

\textit{Proof:} Please, see Appendix \ref{ap1}.  \hfill\qedsymbol \vspace{-3mm}

\subsection{Statistical analysis of channel gains}\label{statch}
In order to proceed with the theoretical analysis, it is crucial to identify the statistical distribution of the gains $\mathcal{\ddot{h}}_{gu} = \max \{1/[\mathbf{H}^{\dagger v}_{gu}(\mathbf{H}^{\dagger v}_{gu})^H]_{gg}, 1/[\mathbf{H}^{\dagger h}_{gu}(\mathbf{H}^{\dagger h}_{gu})^H]_{gg} \}$ and $\mathfrak{X}_{gu} = \left|[\mathbf{H}^{\dagger \ddot{p}}_{gu}\mathbf{\underline{H}}_{gu}^{t \ddot{p}}\mathbf{x}^t]_{g} \right|^2$. This task will be performed in this subsection. By turning our attention to the $u$th user in the $g$th vertical subset, let us identify the distribution of $[\mathbf{H}^{\dagger v}_{gu}(\mathbf{H}^{\dagger v}_{gu})^H]_{gg}$. In particular, by recalling \eqref{eq:ch1} and \eqref{eq:ch24}, the matrix $\mathbf{H}^{\dagger v}_{gu}(\mathbf{H}^{\dagger v}_{gu})^H$ can be expanded as
\begin{align}\label{eq:statch}
        &\mathbf{H}^{\dagger v}_{gu}(\mathbf{H}^{\dagger v}_{gu})^H =  [\mathbf{\Tilde{P}}^H \mathbf{\Tilde{H}}_{gu}^{v v} (\mathbf{\Tilde{H}}_{gu}^{v v})^H\mathbf{\Tilde{P}} ]^{-1} =
        \left( \mathbf{\Tilde{P}}^H 
        \mathbf{U}\mathbf{\Lambda}^{\frac{1}{2}}
        (\mathbf{G}_{gu}^{ vv})^H (\bm{\Phi}_{gu}^{vv})^H \mathbf{S}_{gu}^{ vv} (\mathbf{S}_{gu}^{ vv})^H \bm{\Phi}_{gu}^{vv} \mathbf{G}_{gu}^{ vv}
        \mathbf{\Lambda}^{\frac{1}{2}} \mathbf{U}^H
        \mathbf{\Tilde{P}} \right. \nonumber\\[-2mm]
        &
        +\hspace{-1mm} \left. 
         \mathbf{\Tilde{P}}^H
        \mathbf{U}\hspace{-.5mm}\mathbf{\Lambda}^{\frac{1}{2}}\hspace{-.5mm}
        (\mathbf{G}_{gu}^{ vh})^H \hspace{-.8mm}  (\bm{\Phi}_{gu}^{hv})^H \mathbf{S}_{gu}^{ vv} \hspace{-.5mm} (\mathbf{S}_{gu}^{ vv})^H \hspace{-.5mm} \bm{\Phi}_{gu}^{hv} \mathbf{G}_{gu}^{ vh}
        \mathbf{\Lambda}^{\frac{1}{2}}\hspace{-.5mm} \mathbf{U}^H
        \mathbf{\Tilde{P}} + 
        \mathbf{\Tilde{P}}^H \hspace{-.5mm}
        \mathbf{U}\mathbf{\Lambda}^{\frac{1}{2}}
        \mathbf{D}_{gu}^{ vv} (\mathbf{D}_{gu}^{ vv})^H \hspace{-.8mm}
        \mathbf{\Lambda}^{\frac{1}{2}} \mathbf{U}^H
        \mathbf{\Tilde{P}} \right)^{-1}.
\end{align}

Given that $\mathbf{S}_{gu}^{ pq}$ is a full rank channel matrix, we have that $\mathbb{E}\{ \mathbf{S}_{gu}^{ vv}(\mathbf{S}_{gu}^{ vv})^H\} = \zeta_{gu}^\text{\tiny IRS-U} \mathbf{I}_{L,L}$. Then, by using \eqref{eq:ch0}, the matrix in \eqref{eq:statch} is further simplified as
\begin{align}\label{eq:statch2} 
        \mathbf{H}^{\dagger v}_{gu}(\mathbf{H}^{\dagger v}_{gu})^H &=
        \left( \frac{1}{2} \zeta_{gu}^\text{\tiny BS-IRS} \zeta_{gu}^\text{\tiny IRS-U}
        \mathbf{\Tilde{P}}^H (\mathbf{\bar{G}}_{gu}^{ vv})^H (\bm{\Phi}_{gu}^{vv})^H
        \bm{\Phi}_{gu}^{vv} \mathbf{\bar{G}}_{gu}^{ vv}
        \mathbf{\Tilde{P}}
        \right. \nonumber\\[-2mm]
        &+ \left. \frac{1}{2}
        \zeta_{gu}^\text{\tiny BS-IRS} \zeta_{gu}^\text{\tiny IRS-U} \mathbf{\Tilde{P}}^H
        (\mathbf{\bar{G}}_{gu}^{ vh})^H (\bm{\Phi}_{gu}^{hv})^H \bm{\Phi}_{gu}^{hv} 
          \mathbf{\bar{G}}_{gu}^{ vh}
        \mathbf{\Tilde{P}} +
        \zeta_{gu}^\text{\tiny BS-U} \mathbf{\Tilde{P}}^H
        \mathbf{\bar{D}}_{gu}^{ vv} (\mathbf{\bar{D}}_{gu}^{ vv})^H
        \mathbf{\Tilde{P}} \right)^{-1} \hspace{-2mm}.
\end{align}

As can be observed, the entries of the matrix above will be the result of the inverse of the sum of three independent matrices. Therefore, one could fully characterize $\mathbf{H}^{\dagger v}_{gu}(\mathbf{H}^{\dagger v}_{gu})^H$ by identifying the distributions of the virtual channels $\mathbf{\Tilde{P}}^H\mathbf{\bar{D}}_{gu}^{ vv}$, $\mathbf{\Tilde{P}}^H(\mathbf{\bar{G}}_{gu}^{ vv})^H (\bm{\Phi}_{gu}^{vv})^H$, and  $\mathbf{\Tilde{P}}^H(\mathbf{\bar{G}}_{gu}^{ hv})^H (\bm{\Phi}_{gu}^{hv})^H$. However, since the elements of $\bm{\Phi}_{gu}^{vv}$ and $\bm{\Phi}_{gu}^{hv}$ result from the optimization problem in \eqref{p4}, which change rapidly with the fast fading channels, determining the exact distribution of \eqref{eq:statch2}, with $L$ assuming any value in $\mathbb{N}_{>0}$, becomes a difficult task.

In face of this mathematical challenge, we study next the limiting case with large number of reflecting elements, i.e., $L \rightarrow \infty$, which is also important since it provides a bound to the maximum achievable performance of the proposed system. As one can observe in \eqref{eq:statch2}, the key step to proceed with the analysis is to study the behavior of $(\bm{\Phi}_{gu}^{pq})^H\bm{\Phi}_{gu}^{pq}$ in the large-scale regime of $L$. The following lemma performs this task.

\paragraph*{Lemma II}
If the matrices $\bm{\Phi}_{gu}^{pq}$, $p,q\in \{v,h\}$, are optimized to cancel out co-polar and cross-polar interference, like in \eqref{p1},  when the number of reflecting elements $L$ becomes large, the magnitude of the reflection coefficients becomes arbitrarily small, i.e., $(\bm{\Phi}_{gu}^{pq})^H\bm{\Phi}_{gu}^{pq} \rightarrow \mathbf{0}_{L,L}$ as $L \rightarrow \infty$, $\forall p,q \in \{v,h\}$.

\vspace{2mm}
\textit{Proof:} Please, see Appendix \ref{ap2}. \hfill\qedsymbol

Based on Lemma II, it becomes clear that the channel matrices corresponding to the reflected link BS-IRS-U in \eqref{eq:statch2} will be attenuated with the increase of the number of reflecting elements $L$. Therefore, in the limiting case with $L \rightarrow \infty$, \eqref{eq:statch2} can be approximated by
\begin{align}\label{eq:statch3}
        \mathbf{H}^{\dagger v}_{gu}(\mathbf{H}^{\dagger v}_{gu})^H \approx
        \left( \zeta_{gu}^\text{\tiny BS-U} \mathbf{\Tilde{P}}^H
        \mathbf{\bar{D}}_{gu}^{ vv} (\mathbf{\bar{D}}_{gu}^{ vv})^H
        \mathbf{\Tilde{P}} \right)^{-1},
\end{align}
which can be characterized as follows. First, remember that $\mathbf{\Tilde{P}}$ is an unitary matrix and $\mathbf{\bar{D}}_{gu}^{ vv}$ follows a complex Gaussian distribution. Consequently, the product $\mathbf{\Tilde{P}}^H\mathbf{\bar{D}}_{gu}^{ vv} \in \mathbb{C}^{\frac{\bar{M}}{2} \times \frac{N}{2}}$ will also follow a complex Gaussian distribution. This leads us to conclude that, when $L\rightarrow\infty$, $\mathbf{H}^{\dagger v}_{gu}(\mathbf{H}^{\dagger v}_{gu})^H$ will converge in distribution to an inverse Wishart distribution with $\frac{N}{2}$ degrees of freedom, and covariance matrix given by $\mathbb{E}\{\mathbf{H}^{\dagger v}_{gu}(\mathbf{H}^{\dagger v}_{gu})^H\} = \left(\zeta_{gu}^\text{\tiny BS-U}\mathbf{\Tilde{P}}^H \mathbf{R} \mathbf{\Tilde{P}} \right)^{-1}$, which is a diagonal matrix. Therefore, given the dimensions of $\mathbf{\Tilde{P}}^H\mathbf{\bar{D}}_{gu}^{ vv}$, the channel gain $1/[\mathbf{H}^{\dagger v}_{gu}(\mathbf{H}^{\dagger v}_{gu})^H]_{gg}$ will converge to the Gamma distribution with shape parameter $(N - \bar{M})/2 + 1$ and rate parameter $ (\zeta_{gu}^\text{\tiny BS-U} [\mathbf{\Tilde{P}}^H \mathbf{R} \mathbf{\Tilde{P}}]_{gg})^{-1}$. A similar analysis can be carried out for the effective channel gain observed in the horizontal polarization. However, its corresponding covariance matrix will be also multiplied by the iXPD factor experienced in the link BS-U, i.e., $\mathbb{E}\{\mathbf{H}^{\dagger h}_{gu}(\mathbf{H}^{\dagger h}_{gu})^H\} = \left( \zeta_{gu}^\text{\tiny BS-U} \chi^{\text{\tiny BS-U}} \mathbf{\Tilde{P}}^H \mathbf{R} \mathbf{\Tilde{P}} \right)^{-1}$. Before we continue, for the sake of simplicity, let $\lambda_{gu} =  (\zeta_{gu}^\text{\tiny BS-U} [\mathbf{\Tilde{P}}^H \mathbf{R} \mathbf{\Tilde{P}}]_{gg})^{-1}$ and $\kappa = (N - \bar{M})/2 + 1$. Then, by recalling that the channel coefficients observed in both polarizations are independent, the cumulative distribution function (CDF) for the effective channel gain $\mathcal{\ddot{h}}_{gu}$ can be derived as
\begin{align}
    F_{\mathcal{\ddot{h}}_{gu}}(x) = F_{\mathcal{h}^{v}_{gu}}(x)F_{\mathcal{h}^{h}_{gu}}(x) = \frac{\gamma\left(\kappa, ( \chi^{\text{\tiny BS-U}})^{-1} \lambda_{gu} x \right) \gamma\left(\kappa, \lambda_{gu} x \right)}{\Gamma\left(\kappa \right)^2},
\end{align}
and the respective probability density function (PDF) can be obtained from the derivative of $F_{\mathcal{\ddot{h}}_{gu}}(x)$, resulting in
\begin{align}
    f_{\mathcal{\ddot{h}}_{gu}}(x) &= 
    \frac{ (\lambda_{gu})^{\kappa} x^{\kappa-1} 
    }{\Gamma\left(\kappa \right)^2} \left( e^{-\lambda_{gu} x}\gamma\left(\kappa, (\chi^{\text{\tiny BS-U}})^{-1}\lambda_{gu} x \right) + (\chi^{\text{\tiny BS-U}})^{-\kappa} e^{-(\chi^{\text{\tiny BS-U}})^{-1}\lambda_{gu} x}\gamma\left(\kappa, \lambda_{gu} x \right) \right).
\end{align}

Note that the effective channel gains of users from horizontal subsets will also have an identical distribution as the above.

Another implication of Lemma II is that, for large values of $L$, the magnitude of the reflection coefficients required for cancelling out all polarization interference will be always less than one. Therefore, when $L\rightarrow \infty$, the solution obtained through the optimization problems \eqref{p4} and \eqref{p5} should converge to that obtained via standard unconstrained least squares problem, as in \eqref{lsp}, and, consequently, the polarization interference term in \eqref{sinreq} will be extinguished, i.e., $\lim_{L\rightarrow \infty} \mathfrak{X}_{gu} \rightarrow 0$. The ergodic rates for this limiting case are derived in the next subsection. \vspace{-3mm}

\subsection{Ergodic rates for the large-scale regime of $L$}\label{secrate}
Now, we derive the ergodic rates for users within each polarization subset. Specifically, we consider a scenario in which the users are assisted by IRSs with a large number of reflecting elements. Therefore, as a consequence of Lemma II, the data rates will not be impacted by polarization interference, but only from errors due to imperfect SIC. Under such considerations, a closed-form expression for the ergodic rates is derived in the following proposition.

\paragraph*{Proposition I} When the $u$th user in the $g$th polarization subset is assisted by an IRS with a large number of reflecting elements, i.e., $L\rightarrow \infty$, and considering degradation from imperfect SIC, it will experience the following ergodic rate
\begin{align}\label{ergprop}
    \bar{R}_{gu} &= \frac{ 1
    }{\ln(2)\Gamma\left(\kappa \right)}  
      \Bigg\{ \MeijerG*{1}{3}{3}{2}{1 - \kappa ,1, 1}{1,0}{\frac{\bar{\alpha}_{gu}}{\lambda_{gu}}}  + \MeijerG*{1}{3}{3}{2}{1 - \kappa ,1, 1}{1,0}{\frac{\chi^{\text{\tiny BS-U}} \bar{\alpha}_{gu}}{\lambda_{gu} }} - \MeijerG*{1}{3}{3}{2}{1 - \kappa ,1, 1}{1,0}{\frac{\tilde{\alpha}_{gu}}{\lambda_{gu}}} \nonumber\\
      &  - \MeijerG*{1}{3}{3}{2}{1 - \kappa ,1, 1}{1,0}{\frac{\chi^{\text{\tiny BS-U}} \tilde{\alpha}_{gu} }{\lambda_{gu}}} - \hspace{-1mm}
      \sum_{n=0}^{\kappa-1} \frac{(\chi^{\text{\tiny BS-U}})^{\kappa} + (\chi^{\text{\tiny BS-U}})^{n}}{n!(\chi^{\text{\tiny BS-U}} + 1 )^{\kappa + n}}
      \left[ \MeijerG*{1}{3}{3}{2}{1 - \kappa - n ,1, 1}{1,0}{\frac{\chi^{\text{\tiny BS-U}} \bar{\alpha}_{gu}}{(\chi^{\text{\tiny BS-U}} + 1)\lambda_{gu}}} \right. \nonumber\\
    & \left. - \MeijerG*{1}{3}{3}{2}{1 - \kappa - n ,1, 1}{1,0}{\frac{\chi^{\text{\tiny BS-U}} \tilde{\alpha}_{gu}}{(\chi^{\text{\tiny BS-U}} + 1)\lambda_{gu}}} \right] \Bigg\}, \qquad 1 \leq u \leq U,
\end{align}
where $ \bar{\alpha}_{gu} = \rho (\alpha_{gu}^2 + \mathfrak{I}_{gu})$, and $ \tilde{\alpha}_{gu} = \rho \mathfrak{I}_{gu}$.

\textit{Proof:} Please, see Appendix \ref{ap3}. \hfill\qedsymbol

Even though \eqref{ergprop} may look complex to interpret, by knowing that all terms with Meijer's G-functions are increasing functions of the SNR $\rho$, which have been numerically verified, we can still extract some insights. First, note that the terms that are functions of $\tilde{\alpha}_{gu}$, which accounts only for interference, are negative. This suggests that such terms are expected to degrade the ergodic rates of the users as long as they experience some interference. On the other hand, the positive term that depends on $\chi^{\text{\tiny BS-U}}$ and $\bar{\alpha}_{gu}$ indicates that the cross-polar transmissions will improve the rate performance of the users. This behavior is indeed expected since the IRSs enables polarization diversity by recycling cross-polar transmissions. \vspace{-2mm}

\section{Simulation Results and Discussions}
In this section, by presenting representative numerical simulation examples, we validate the theoretical analysis carried out in the last section and demonstrate the potential performance gains that the proposed dual-polarized IRS-MIMO-NOMA scheme can achieve over conventional systems. Specifically, we use as baseline schemes the classical MIMO-OMA system, where users are served via time division multiple access, and the conventional single-polarized and dual-polarized MIMO-NOMA systems, whose implementation details can be found in \cite{ni3}.

For a fair performance comparison, in both single and dual-polarized schemes, we employ at the BS a linear array with $M=90$ transmit antennas. However, as explained in the System Model Section, the antenna elements in the dual-polarized systems are arranged into co-located pairs, thereby, resulting in $\frac{M}{2} = 45$ pairs of dual-polarized antennas. For modeling the scattering environment and the correlation between transmit antennas, we generate the covariance matrices in \eqref{covmat1} and \eqref{covmat2} through the one-ring geometrical model \cite{ni3,ref6,ref1}, where we consider the existence of $K = 4$ spatial clusters, each with $30$~m of radius and located at $120$~m from the BS. In addition, the BS's antenna array is directed to the first cluster that is positioned at the azimuth angle of $30^\circ$. This is the cluster from which the simulation results are generated, which comprises $G=\bar{M}=4$ groups, each one containing $U=4$ users. In particular, we focus on the first group, where the users $1$, $2$, $3$ and $4$ are located, respectively, at $135$~m, $125$~m, $115$~m, and $105$~m from the BS. A fixed power allocation is adopted, in which we set $\alpha_{1}^2 = 0.4, \alpha_{2}^2 =0.35, \alpha_{3}^2 = 0.2, \alpha_{4}^2 = 0.05$. %, i.e., more power is allocated to distant users and less power to the closer ones.
Moreover, we assume that the distances from the BS to each IRS are the same as that from the BS to its connected user. Under these assumptions, the fading coefficients for the links BS-U and BS-IRS are configured as $\zeta_{u}^\text{\tiny BS-U} = \zeta_{u}^\text{\tiny BS-IRS} = \varrho d_{u}^{-\eta}$, where $d_{u}$ is the distance between the BS and the $u$th user and its serving IRS, $\varrho = 2\times 10^4$ is an array gain parameter that is configured at the BS according to the desired receivers' performance \cite{SenaISIC2020}, and $\eta = 2$ is the path-loss exponent. Regarding the link IRS-U, since an IRS is a passive device, we discard the array gain and model the corresponding fading coefficient as $\zeta_{gu}^\text{\tiny IRS-U} = \tilde{d}^{-\eta}$, where $\tilde{d} = 20$~m for all IRSs, i.e., users are positioned $20$~m apart from its serving IRS. Other parameters that have not been mentioned will assume different values throughout the simulation examples. Those will be informed accordingly next.

\begin{figure}[t]
	\includegraphics[scale=0.5]{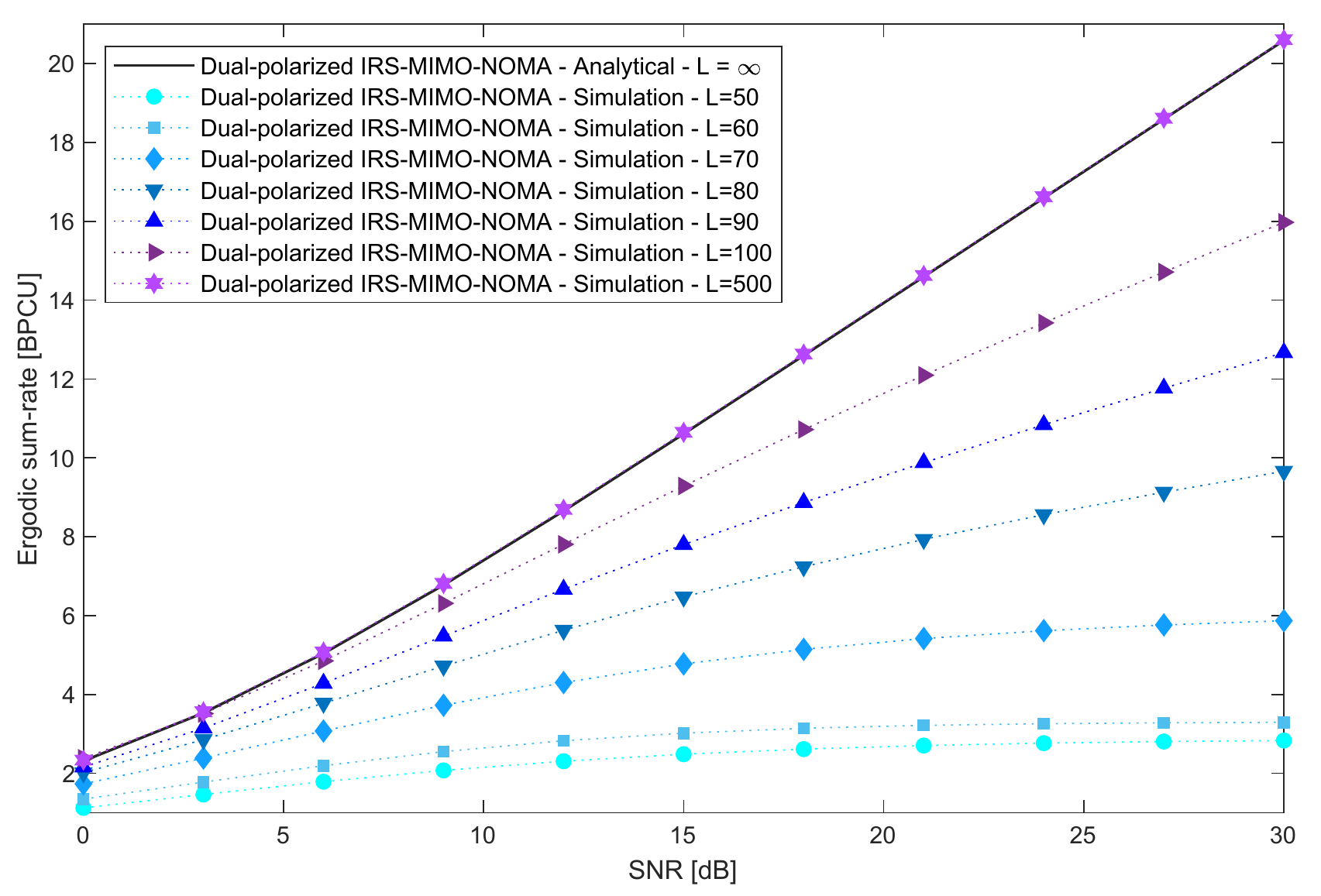}
	\centering
	\caption{Simulated and analytical ergodic sum-rates considering perfect SIC. Effect of the increase in the number of dual-polarized reflecting elements ($N = 4, \chi = 0.5, \xi = 0$).}\label{f1}
\end{figure}

Fig. \ref{f1} brings the simulated and analytical ergodic sum-rate curves, generated by $\sum_{i=1}^U \bar{R}_{gi}$, for various values of dual-polarized reflecting elements $L$, and considering perfect SIC decoding. As one can see, for small numbers of reflecting elements, when optimizing the IRSs through \eqref{p4} and \eqref{p5}, the simulated ergodic sum-rate curves reach values lower than that from the analytical curve obtained by solving \eqref{ergprop}. This behavior is explained by the fact that the IRS cannot eliminate all polarization interference when the number of reflecting elements is small, which degrades the system sum-rate. However, as the number of reflecting elements increases, the polarization interference decreases, and the sum-rate improves, approaching the analytical one. For instance, when $500$ dual-polarized elements are considered, the simulated sum-rate matches perfectly the analytical curve. Such performance is in total agreement with Lemma II and the analytical derivation of Section \ref{secrate}, therefore, providing the first validation to our analysis.

\begin{figure}[t]
	\includegraphics[scale=0.5]{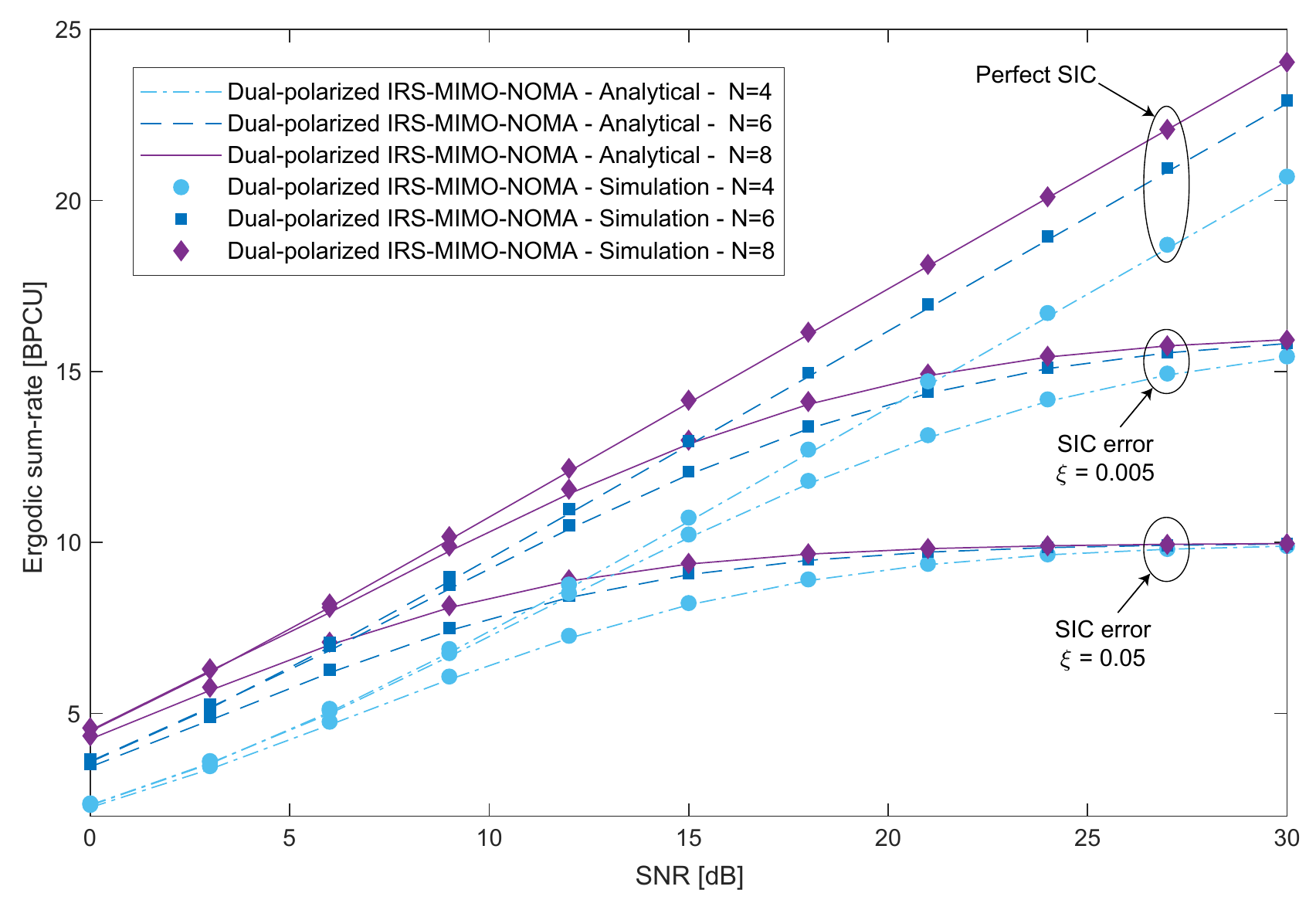}
	\centering
	\caption{Simulated and analytical ergodic sum-rates for various levels of SIC error propagation ($L = 500, \chi = 0.5$).}\label{f2}
\end{figure}

\begin{figure}[t]
	\includegraphics[scale=0.5]{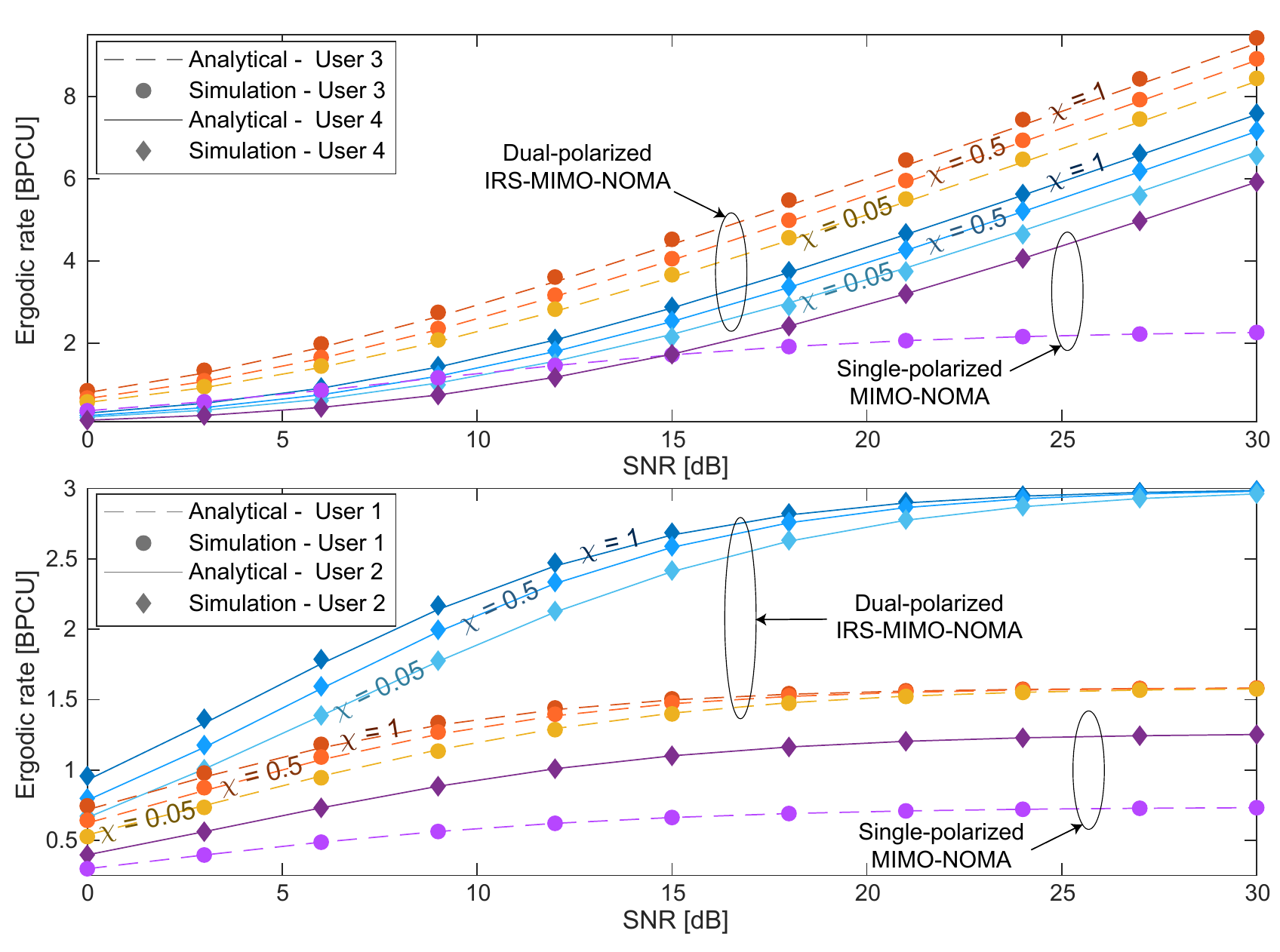}
	\centering
	\caption{Simulated and analytical ergodic rates for different values of iXPD ($L = 500, N = 4, \xi= 0$).}\label{f3}
\end{figure}

To further corroborate the analysis for large values of $L$, we present in Fig. \ref{f2} the sum-rates and in Fig. \ref{f3} the individual ergodic rates for $500$ reflecting elements, in which, in all considered cases, a perfect agreement between simulated and analytical curves can be observed. Specifically, Fig. \ref{f2} shows the effects of SIC error propagation on the system performance for different numbers of receive antennas. As one can notice, when the users face imperfect SIC, their sum-rate curves become limited to a saturation point that deteriorates with the increase of the error factor $\xi$. This happens due to the fact that all users, even the strongest one, experience interference when $\xi \neq 0$, thereby, leading to the observed limited performance. Such behavior confirms the insights raised in the last paragraph of Section \ref{secrate}.

Fig. \ref{f3} depicts the impact of the level of cross-polar transmissions in the users' ergodic rates considering perfect SIC, in which results for different values for the iXPD parameter $\chi$ are shown. In addition to validating the theoretical analysis, this figure shows how beneficial the proposed scheme can be to improve the performance of each user. It also becomes clear that, with the help of IRSs, depolarization phenomena can be transformed into an advantage, e.g., the higher the iXPD, the greater the performance gains. For instance, in the conventional single-polarized system, when the SNR is $30$~dB, the rate of user $3$ is limited to only $2.27$ bits per channel use (BPCU). On the other hand, when this same user is served via the IRS-MIMO-NOMA scheme, for a low iXPD of $\chi = 0.05$, and an SNR of $30$~dB, its rate can reach $8.44$~BPCU, which is more than three times greater than that achieved in the single-polarized scheme. When we consider a high iXPD of $\chi = 1$, the achievable ergodic rate of the user $3$ becomes even more remarkable, reaching up to $9.42$~BPCU. Impressive performance gains can be also observed in all the other users, with their rates remarkably outperforming those achievable in the conventional single-polarized scheme. These improvements are mainly due to two features of the proposed IRS-MIMO-NOMA system, already explained in previous sections. That is, firstly, the IRSs enable the users to exploit polarization diversity, and, secondly, the users are able to perform SIC considering interference only from their own polarization subset. Therefore, in addition to benefit from diversity, users in the IRS-MIMO-NOMA system are impacted by less interference than they are in the conventional MIMO-NOMA counterpart.

\begin{figure}[t]
	\includegraphics[scale=0.5]{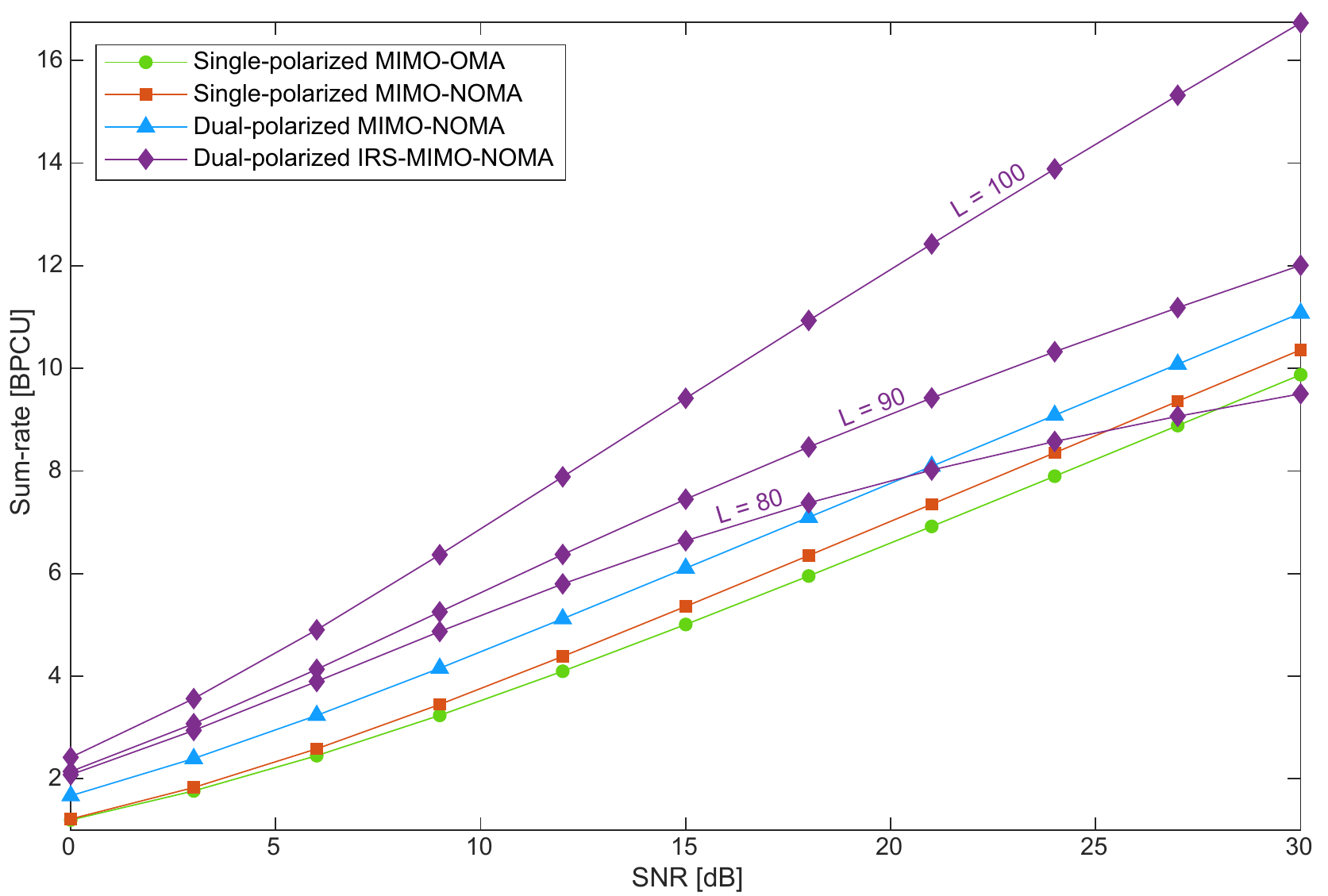}
	\centering
	\caption{Simulated ergodic sum-rates. Comparison between proposed dual-polarized IRS-MIMO-NOMA and conventional schemes ($N = 4, \chi = 0.5, \xi= 0$).}\label{f5}
\end{figure}

\begin{figure}[t]
	\includegraphics[scale=0.5]{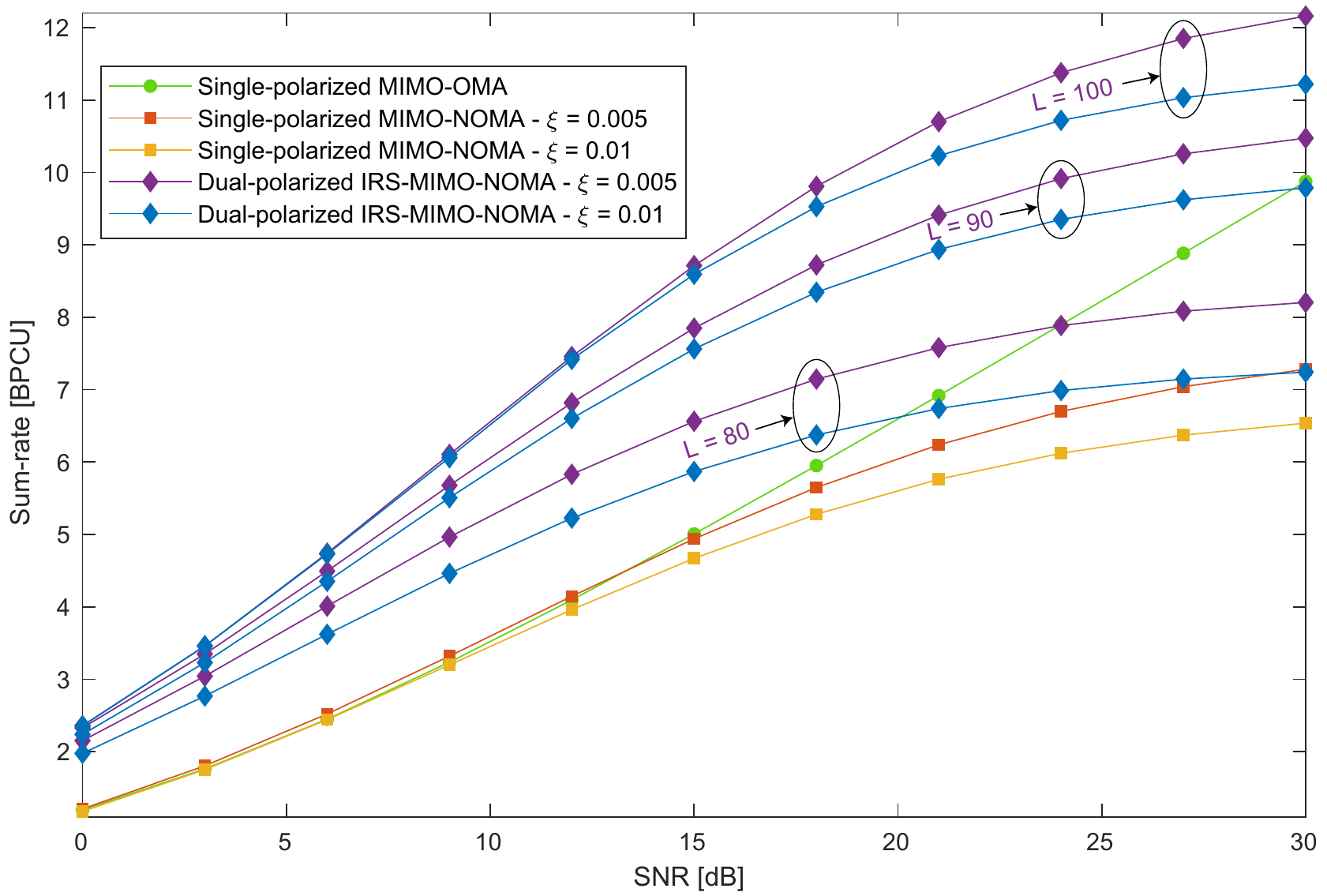}
	\centering
	\caption{Simulated ergodic sum-rates. Comparison between proposed dual-polarized IRS-MIMO-NOMA and conventional schemes under imperfect SIC ($N = 4, \chi = 0.5$).}\label{f6}
\end{figure}

In Fig. \ref{f5}, we compare the sum-rate performance of the proposed IRS-MIMO-NOMA scheme and other conventional systems assuming perfect SIC. As one can notice, when $L=80$, from $21$~dB onward, the proposed scheme is outperformed by the dual-polarized MIMO-NOMA counterpart, and when the SNR reaches $30$~dB, the MIMO-OMA system is the one that achieves the best performance. However, with a slight increase in the number of reflecting elements, from $L = 80$ to $L = 90$, the IRS-MIMO-NOMA scheme can already outperform all the other baseline schemes, in all considered SNR range.
%Specifically, when the SNR is $24$~dB, the IRS-MIMO-NOMA system with $L=90$ reflecting elements achieves a sum-rate of $10.42$~BPCU, which represents a gain of $1.24$~BPCU and $1.97$~BPCU over, respectively, the dual-polarized and single-polarized MIMO-NOMA systems, and $2.43$~BPCU over the conventional MIMO-OMA counterpart. By increasing the number of reflecting elements to $L=100$, the performance gains become even higher, resulting in an impressive gap of $4.81$~BPCU between the sum-rates of the IRS-MIMO-NOMA and the dual-polarized MIMO-NOMA schemes.
Finally, Fig. \ref{f6} shows how well the dual-polarized IRS-MIMO-NOMA system performs in comparison with the single-polarized MIMO-OMA and MIMO-NOMA counterparts in the presence of SIC error propagation. As can be seen, even though the sum-rate of all NOMA-based schemes are caped in the high-SNR regime, the proposed IRS-MIMO-NOMA system is significantly more robust to SIC errors than the conventional single-polarized MIMO-NOMA. For instance, for a SIC error factor of $\xi = 0.005$, the single-polarized MIMO-NOMA can only slightly outperform the MIMO-OMA scheme for SNR values lower than $15$~dB. When the error is $\xi = 0.01$, the sum-rate degradation becomes so severe that, in the whole SNR range, the MIMO-OMA system outperforms the MIMO-NOMA counterpart. On the other hand, even when considering $L=80$ reflecting elements, and an error of $\xi = 0.01$, the IRS-MIMO-NOMA can reach sum-rates remarkably higher than those achieved by the conventional schemes, being outperformed by the MIMO-OMA scheme only in SNR values above $20$~dB. Moreover, for $L=90$ and $L=100$, the IRS-MIMO-NOMA scheme always achieves the best performance. For example, when $\xi = 0.005$, $L=100$ and the SNR is $18$~dB, the IRS-MIMO-NOMA scheme reaches an expressive sum-rate of $9.81$~BPCU, which is an increase of $3.86$~BPCU over the MIMO-OMA system and $4.17$~BPCU over the single-polarized MIMO-NOMA. \vspace{-2mm}

\section{Conclusions}
In this work, by exploiting the capabilities of dual-polarized IRSs, we proposed and investigated a novel strategy for improving the performance of dual-polarized massive MIMO-NOMA networks under the impact of imperfect SIC. The detailed construction of the beamforming and reception matrices was provided, and an efficient procedure for optimizing the IRS reflecting elements was developed. Moreover, we carried out an insightful mathematical analysis, in which the ergodic rates for large numbers of reflecting elements were derived. Our numerical results revealed that the proposed dual-polarized IRS-MIMO-NOMA scheme can achieve remarkable performance gains over conventional single-polarized and dual-polarized systems and that cross-polar transmissions can further improve the ergodic rates of the users. \vspace{-2mm}

\appendices

\section{Proof of Lemma I}\label{ap1}
\renewcommand{\theequation}{A-\arabic{equation}}
\setcounter{equation}{0}

Given the data symbol in \eqref{sigbsic4}, when the $u$th user in the polarization subset $\mathcal{U}^{p}$, $p\in \{v,h\}$, of the $g$th group decodes the message intended to the $i$th user, $\min\{\mathcal{U}^{p}\} < i<u$, $i \in \mathcal{U}^{p}$, it experiences the following SINR
\begin{align}\label{eqB2}
\gamma_{gu}^{i} &= |\alpha_{gi}x_{gi}|^2 \left( \sum_{m \in \{a | \hspace{.5mm} a > i, \hspace{.5mm} a\in \mathcal{U}^{p} \}} \hspace{-8mm} |\alpha_{gm}x_{gm}|^2 +
\xi \hspace{-8mm} \sum_{n \in \{b | \hspace{.5mm} b < i, \hspace{.5mm} b\in \mathcal{U}^{p} \}} \hspace{-8mm} |\alpha_{gn}x_{gn}|^2 +
\left|[\mathbf{H}^{\dagger \ddot{p}}_{gu}\mathbf{\underline{H}}_{gu}^{t \ddot{p}}\mathbf{x}^t]_{g} \right|^2 + 
| [\mathbf{H}^{\dagger \ddot{p}}_{gu}\mathbf{n}^{\ddot{p}}_{gu}]_{g}|^2 \right)^{-1} \nonumber\\
&= \alpha_{gi}^2 \left( \sum_{m \in \{a | \hspace{.5mm} a > i, \hspace{.5mm} a\in \mathcal{U}^{p} \}} \hspace{-8mm} \alpha_{gm}^2 +
\hspace{4mm} \xi \hspace{-8mm} \sum_{n \in \{b | \hspace{.5mm} b < i, \hspace{.5mm} b\in \mathcal{U}^{p} \}} \hspace{-8mm} \alpha_{gn}^2 +
\left|[\mathbf{H}^{\dagger \ddot{p}}_{gu}\mathbf{\underline{H}}_{gu}^{t \ddot{p}}\mathbf{x}^t]_{g} \right|^2 + 
\sigma_n^2 [\mathbf{H}^{\dagger \ddot{p}}_{gu}(\mathbf{H}^{\dagger \ddot{p}}_{gu})^H]_{gg} \right)^{-1}.
\end{align}

By defining $\rho = 1/\sigma^2_n$ as the SNR, and denoting the effective channel gain by $\mathcal{\ddot{h}}_{gu} = \max\{\mathcal{h}^{v}_{gu}, \mathcal{h}^{h}_{gu}\}=\max \{1/[\mathbf{H}^{\dagger v}_{gu}(\mathbf{H}^{\dagger v}_{gu})^H]_{gg}, 1/[\mathbf{H}^{\dagger h}_{gu}(\mathbf{H}^{\dagger h}_{gu})^H]_{gg} \}$, the SINR can be rewritten as
\begin{align}\label{eqB3}
\gamma_{gu}^{i} &= \rho\mathcal{\ddot{h}}_{gu} \alpha_{gi}^2 
\left[
\rho\mathcal{\ddot{h}}_{gu} \left( \sum_{m=i+1}^{\max\{\mathcal{U}^{p}\}} \hspace{-2mm} \alpha_{gm}^2 
+ \xi \hspace{-4mm}\sum_{n=\min\{\mathcal{U}^{p}\}}^{i-1} \hspace{-4mm} \alpha_{gn}^2 
+ \left|[\mathbf{H}^{\dagger \ddot{p}}_{gu}\mathbf{\underline{H}}_{gu}^{t \ddot{p}}\mathbf{x}^t]_{g} \right|^2 \right) + 
1\right]^{-1}.
\end{align}

Note that when the weakest user, i.e., the $1$st user corresponding to $\min\{\mathcal{U}^{p}\}$, detects its symbol, it will experience interference from everyone else, but it will not face imperfect SIC. On the other hand, when the user with the best channel gain, i.e., the user corresponding to the maximum index in $\mathcal{U}^{p}$, decodes its symbol, there will be no interference from higher-order users, but only from imperfect SIC.
%Also, note that the polarization interference term originated at the BS is independent of SIC, i.e., it depends only on the effectiveness of IRS for canceling undesired transmissions.
Under these observations, we denote the polarization interference by $\mathfrak{X}_{gu} = \left|[\mathbf{H}^{\dagger \ddot{p}}_{gu}\mathbf{\underline{H}}_{gu}^{t \ddot{p}}\mathbf{x}^t]_{g} \right|^2$, and the total SIC interference by 
\begin{align}
    \mathfrak{I}_{gi} &=
    \begin{cases}
        \sum_{m=i+1}^{\max\{\mathcal{U}^{p}\}} \alpha_{gm}^2, \hspace{-2mm} & \text{ if } i = \min\{\mathcal{U}^{p}\},\\
        \sum_{m=i+1}^{\max\{\mathcal{U}^{p}\}} \alpha_{gm}^2 + \xi \sum_{n=\min\{\mathcal{U}^{p}\}}^{i-1} \alpha_{gn}^2, \hspace{-2mm}& \text{ if } \min\{\mathcal{U}^{p}\} < i \leq u < \max\{\mathcal{U}^{p}\},\\
        \xi \sum_{n=1}^{i-1} \alpha_{gn}^2, \hspace{-2mm} & \text{ if } i = u = \max\{\mathcal{U}^{p}\}.
    \end{cases}
\end{align}

Then, by applying the above definitions in \eqref{eqB3}, we can achieve the final SINR expression, as in \eqref{sinreq}, which completes the proof. \hfill\qedsymbol

\section{Proof of Lemma II}\label{ap2}
\renewcommand{\theequation}{B-\arabic{equation}}
\setcounter{equation}{0}

First, note that $(\bm{\Phi}_{gu}^{pq})^H\bm{\Phi}_{gu}^{pq}$ is a diagonal matrix whose entries are the squared magnitude of the reflection coefficients, i.e., $(\omega_{gu,l}^{pq})^2$. Therefore, we aim to investigate the behavior of $(\omega_{gu,l}^{pq})^2$ when $L\rightarrow \infty$. For this, let us start by relaxing the unity $\mathpzc{L}_\infty$ norm constraint \eqref{p2b}, and rewriting the problem in \eqref{p2} as\vspace{-2mm}
\begin{align}\label{lsp}
        &\underset{\bm{\theta}^{vv}_{gu}, \bm{\theta}^{hv}_{gu}}{\min}
         \left\|\setlength{\arraycolsep}{1pt}
       \mathbf{\bar{K}}_{gu} \left[(\bm{\theta}^{vv}_{gu})^T, (\bm{\theta}^{hv}_{gu})^T\right]^T \hspace{-2mm}  +
    \mathbf{d}^{hv}_{gu}
        \right\|^2,
\end{align}
which consists of a standard least squares problem that, by assuming $L \geq N r_k^\star$, has optimal solution given by\vspace{-2mm}
\begin{align}\label{optsol}
    \left[(\bm{\Dot{\theta}}^{vv}_{gu})^T, (\bm{\Dot{\theta}}^{hv}_{gu})^T\right]^T  = \mathbf{\bar{K}}_{gu}^H(\mathbf{\bar{K}}_{gu}\mathbf{\bar{K}}_{gu}^H)^{-1} \mathbf{d}^{hv}_{gu},
\end{align}
which is the solution with minimum $\mathpzc{L}_2$ norm. Then, it follows that $\mathbf{\bar{K}}_{gu}\left[(\bm{\Dot{\theta}}^{vv}_{gu})^T, (\bm{\Dot{\theta}}^{hv}_{gu})^T\right]^T + \mathbf{d}^{hv}_{gu} = 0$, which implies\vspace{-2mm}
\begin{align}\label{impp}
   \sum_{l=1}^{2L} [\mathbf{\bar{K}}_{gu}]_{il} \renewcommand*{\arraystretch}{1}
   \begin{bmatrix}
          \bm{\Dot{\theta}}^{vv}_{gu} \\
          \bm{\Dot{\theta}}^{hv}_{gu}
   \end{bmatrix}_{l} = -\left[\mathbf{d}^{hv}_{gu}\right]_{i}, \forall i = 1, \cdots, \frac{N}{2} r_k^\star.
\end{align}
Recall that $\left[\mathbf{d}^{hv}_{gu}\right]_{i}$ is a complex Gaussian random variable with zero mean and unit variance.
Consequently, the sum on the left-hand side of \eqref{impp} will also have zero mean and unity variance $\forall L \geq N r_k^\star \in \mathbb{N}_{>0}$.
Therefore, we can exploit the independence of $[\mathbf{\bar{K}}_{gu}]_{il}$ and $\left[\left[(\bm{\Dot{\theta}}^{vv}_{gu})^T, (\bm{\Dot{\theta}}^{hv}_{gu})^T\right]^T\right]_l$ and write
\begin{align}\label{impp2}
   \sum_{l=1}^{2L} \mathbb{E}\left\{\left| [\mathbf{\bar{K}}_{gu}]_{il} \right|^2 \right\}
   \mathbb{E} \left\{\left|\renewcommand*{\arraystretch}{1}
   \begin{bmatrix}
          \bm{\Dot{\theta}}^{vv}_{gu} \\
          \bm{\Dot{\theta}}^{hv}_{gu}
   \end{bmatrix}_{l} \right|^2\right\} = \mathbb{E} \left\{ \left|\left[\mathbf{d}^{hv}_{gu}\right]_{i}\right|^2 \right\} = 1.
\end{align}

As long as the reflection coefficients are optimized based on \eqref{optsol}, the sum in \eqref{impp2} will always converge to $1$, independently of $L$. By knowing this beforehand, we need to check the convergence behavior of each term of the above sum separately. First, recall that the entries of $\mathbf{\bar{K}}_{gu}$ also result from independent complex Gaussian random variables with unity variance. Because of this, we have that $\lim_{L\rightarrow\infty}\sum_{l=1}^{2L} \mathbb{E}\{|[\mathbf{\bar{K}}_{gu}]_{il}|^2\} \rightarrow \infty$. Therefore, the sum in \eqref{impp2} will only converge if $\left|\left[\left[(\bm{\Dot{\theta}}^{vv}_{gu})^T, (\bm{\Dot{\theta}}^{hv}_{gu})^T\right]^T\right]_l\right|^2 =  (\omega_{gu,l}^{pq})^2 \rightarrow 0$, $\forall l = 1, \cdots, 2L$, and we can conclude that $(\bm{\Phi}^{pq}_{gu})^H\bm{\Phi}^{pq}_{gu} \rightarrow \mathbf{0}_{L,L}, \text{ as } L \rightarrow \infty, \forall p,q \in \{v,h\},$, which completes the proof. \hfill\qedsymbol

\section{Proof of Proposition I}\label{ap3}
\renewcommand{\theequation}{C-\arabic{equation}}
\setcounter{equation}{0}
%As demonstrated in Section \ref{statch}, for $L$ large enough, all transmissions intended for interfering polarization subsets can be successfully eliminated. In this limiting scenario, the only interference left will be that resulted from the SIC decoding process, which consist of errors from imperfect SIC and interference from stronger users. More specifically, 
By relying on Lemma II, when $L\rightarrow \infty$, %the SINR expression proposed in Lemma I can be simplified, and
% \begin{align}\label{sinrsimp}
% \gamma_{gu}^{i} &= \frac{\rho\mathcal{\ddot{h}}_{gu} \alpha_{gi}^2}{\rho\mathcal{\ddot{h}}_{gu}\mathfrak{I}_{gi} +  
% 1}, & 1 \leq i \leq u \leq U.
% \end{align}
% As a result, the 
the $u$th user experiences the following data rate
\begin{align}\label{instrate}
R_{gu} %&= \log_2 \left( 1 + \frac{\rho\mathcal{\ddot{h}}_{gu} \alpha_{gu}^2}{\rho\mathcal{\ddot{h}}_{gu}\mathfrak{I}_{gu} + 1} \right) = \log_2 \left(\frac{\rho\mathcal{\ddot{h}}_{gu}(\alpha_{gu}^2 + \mathfrak{I}_{gu}) + 1}{\rho\mathcal{\ddot{h}}_{gu}\mathfrak{I}_{gu} + 1} \right) \nonumber\\
&= \log_2\left(\rho\mathcal{\ddot{h}}_{gu}(\alpha_{gu}^2 + \mathfrak{I}_{gu}) + 1\right) - \log_2\left(\rho\mathcal{\ddot{h}}_{gu}\mathfrak{I}_{gu} + 1\right), \qquad 1 \leq u \leq U.
\end{align}

The ergodic rate can be then derived from the expectation of $R_{gu}$, i.e.,
\begin{align}\label{ergp1}
\bar{R}_{gu} &= \int_{0}^{\infty} \left[ \log_2\left( \rho (\alpha_{gu}^2 + \mathfrak{I}_{gu})x + 1\right) - \log_2\left( \rho \mathfrak{I}_{gu}x + 1\right) \right] f_{\mathcal{\ddot{h}}_{gu}}(x) dx.
\end{align}

Next, by denoting $ \bar{\alpha}_{gu} = \rho (\alpha_{gu}^2 + \mathfrak{I}_{gu})$ and $ \tilde{\alpha}_{gu} = \rho \mathfrak{I}_{gu}$, and replacing the PDF of $\mathcal{\ddot{h}}_{gu}$ in \eqref{ergp1}, we obtain
\begin{align}\label{ergpI}
    \bar{R}_{gu} &= \frac{ (\lambda_{gu})^{\kappa} 
    }{\Gamma\left(\kappa \right)^2}
    \int_{0}^{\infty} \left[\log_2\left(\bar{\alpha}_{gu}x + 1 \right) - \log_2\left(\tilde{\alpha}_{gu}x + 1 \right)\right] x^{\kappa-1} e^{-\lambda_{gu} x}\gamma\left(\kappa, (\chi^{\text{\tiny BS-U}})^{-1}\lambda_{gu} x \right) dx \nonumber\\
    &
    + \frac{ (\lambda_{gu})^{\kappa} 
    }{\Gamma\left(\kappa \right)^2 (\chi^{\text{\tiny BS-U}})^{\kappa}}
    \int_{0}^{\infty} \left[\log_2\left(\bar{\alpha}_{gu}x + 1 \right) - \log_2\left(\tilde{\alpha}_{gu}x + 1 \right)\right] x^{\kappa-1} e^{-(\chi^{\text{\tiny BS-U}})^{-1}\lambda_{gu} x}\gamma\left(\kappa, \lambda_{gu} x \right) dx
    \nonumber\\
    & \triangleq I_{1} + I_{2}.
\end{align}

First, let us focus on solving $I_{1}$. By applying the Meijer's G-function representation for $\ln(x + 1)$ \cite[eq. (2.6.6)]{Mathai73} and exploiting the series representation of the incomplete gamma function in \cite[eq. (8.352.6)]{ref8}, we can rewrite $I_{1}$ as follows
\begin{align}\label{ergp2}
    &I_{1} = \frac{ (\lambda_{gu})^{\kappa} 
    }{\ln(2)\Gamma\left(\kappa \right)} \left[ \int_{0}^{\infty} \left[ \MeijerG*{1}{2}{2}{2}{1, 1}{1,0}{\bar{\alpha}_{gu} x}
    - \MeijerG*{1}{2}{2}{2}{1, 1}{1,0}{\tilde{\alpha}_{gu} x} \right] x^{\kappa-1} e^{-\lambda_{gu}x} dx \right.
    \nonumber\\
    &\left. - \hspace{-1mm} \sum_{n=0}^{\kappa-1} \frac{(\lambda_{gu})^n}{n!(\chi^{\text{\tiny BS-U}})^{n}} \int_{0}^{\infty} \left[ \MeijerG*{1}{2}{2}{2}{1, 1}{1,0}{\bar{\alpha}_{gu} x}
    - \MeijerG*{1}{2}{2}{2}{1, 1}{1,0}{\tilde{\alpha}_{gu} x} \right] x^{\kappa + n - 1} e^{-\lambda_{gu}(1 + (\chi^{\text{\tiny BS-U}})^{-1}) x} dx \right].
\end{align}

Then, by exploiting the Laplace transform property for Meijer's G-functions \cite[eq. (5.6.3.1)]{Keke69}, and performing some manipulations in \eqref{ergp2}, $I_{1}$ can be derived as
\begin{align}\label{pfact1}
    I_{1} &= \frac{ 1
    }{\ln(2)\Gamma\left(\kappa \right)}  
      \Bigg\{ \MeijerG*{1}{3}{3}{2}{1 - \kappa ,1, 1}{1,0}{\frac{\bar{\alpha}_{gu}}{\lambda_{gu}}} - \MeijerG*{1}{3}{3}{2}{1 - \kappa ,1, 1}{1,0}{\frac{\tilde{\alpha}_{gu}}{\lambda_{gu}}} - \hspace{-1mm} \sum_{n=0}^{\kappa-1} \frac{(1 + (\chi^{\text{\tiny BS-U}})^{-1})^{-\kappa - n}}{n!(\chi^{\text{\tiny BS-U}})^{n}} \nonumber\\
    &\times \left[ \MeijerG*{1}{3}{3}{2}{1 - \kappa - n ,1, 1}{1,0}{\frac{\bar{\alpha}_{gu}}{\lambda_{gu}(1 + (\chi^{\text{\tiny BS-U}})^{-1})}} - \MeijerG*{1}{3}{3}{2}{1 - \kappa - n ,1, 1}{1,0}{\frac{\tilde{\alpha}_{gu}}{\lambda_{gu}(1 + (\chi^{\text{\tiny BS-U}})^{-1})}} \right] \Bigg\}.
\end{align}
A similar analysis can be carried out to solve $I_{2}$, which is not shown here due to space constraints. 
% \begin{align}\label{pfact2}
%     I_{2} &= \frac{ 1
%     }{\ln(2)\Gamma\left(\kappa \right)}  
%       \Bigg\{ \MeijerG*{1}{3}{3}{2}{1 - \kappa ,1, 1}{1,0}{\frac{\bar{\alpha}_{gu}}{\lambda_{gu} (\chi^{\text{\tiny BS-U}})^{-1} }} - \MeijerG*{1}{3}{3}{2}{1 - \kappa ,1, 1}{1,0}{\frac{\tilde{\alpha}_{gu}}{\lambda_{gu} (\chi^{\text{\tiny BS-U}})^{-1}}} \nonumber\\
%     &- \hspace{-1mm} \sum_{n=0}^{\kappa-1} \frac{(1 + (\chi^{\text{\tiny BS-U}})^{-1})^{-\kappa - n}}{n!(\chi^{\text{\tiny BS-U}})^{k}}  \left[ \MeijerG*{1}{3}{3}{2}{1 - \kappa - n ,1, 1}{1,0}{\frac{\bar{\alpha}_{gu}}{\lambda_{gu}(1 + (\chi^{\text{\tiny BS-U}})^{-1})}} \right. \nonumber\\
%     & \left. - \MeijerG*{1}{3}{3}{2}{1 - \kappa - n ,1, 1}{1,0}{\frac{\tilde{\alpha}_{gu}}{\lambda_{gu}(1 + (\chi^{\text{\tiny BS-U}})^{-1})}} \right] \Bigg\}.
% \end{align}
Then, after replacing $I_{1}$ and $I_{2}$ in \eqref{ergpI}, and performing some manipulations, the final ergodic rate expression can be obtained as in \eqref{ergprop}, which completes the proof. \hfill\qedsymbol\vspace{-2mm}

%...

\ifCLASSOPTIONcaptionsoff
\newpage
\fi

%\newpage

\bibliographystyle{IEEEtran}
\bibliography{bibtex/references}

\end{document}